\documentclass[12pt]{article}
\usepackage{amsmath}
\usepackage{amssymb}
\usepackage{epsfig}
\usepackage{amsthm}
\usepackage{stmaryrd}
\usepackage{bm}
\usepackage{hyperref}

\renewcommand{\arraystretch}{1.5}

\numberwithin{equation}{section}
\numberwithin{figure}{section}

\def\eq#1{(\ref{eq:#1})}
\def\lineup{\!\!\!\!\!\!\!\!&&}

\newcommand{\Tr}{\mathop{\rm Tr}\nolimits}

\def\d{\partial}
\def\eps{\epsilon}

\def\fraction#1#2{ { \textstyle \frac{#1}{#2} }}
\def\half{\fraction{1}{2}}

\newtheorem*{``theorem''}{Theorem}

\begin{document}

\begin{titlepage}

\begin{center}

\vskip 1.0cm {\large \bf{Analytic Solution for Tachyon Condensation }}\\ 
\vskip .2cm
{\large \bf{in Berkovits' Open Superstring Field Theory}}\\
\vskip 1.5cm
{\large Theodore Erler\footnote{Email: tchovi@gmail.com}}

\vskip 1.0cm

{\it {Institute of Physics of the ASCR, v.v.i.} \\
{Na Slovance 2, 182 21 Prague 8, Czech Republic}}\\
\vskip 1.5cm

{\bf Abstract}

\end{center}

We present an analytic solution for tachyon condensation on a 
non-BPS D-brane in Berkovits' open superstring field theory. 
The solution is presented as a product of $2\times 2$ matrices 
in two distinct $GL_2$ subgroups of the open string star algebra. All string
fields needed for computation of the nonpolynomial action can be derived 
in closed form, and the action produces the expected non-BPS D-brane 
tension in accordance with Sen's conjecture. We also comment on how D-brane 
charges may be encoded in the topology of the tachyon vacuum gauge orbit.

\noindent

\noindent
\medskip

\end{titlepage}

\tableofcontents

\section{Introduction}

Tachyon condensation on unstable D-branes has always been a challenging problem
to study, both because the phenomenon is intrinsically ``stringy''---the 
tachyon $\mathrm{mass}^2$ is proportional to $\frac{1}{\alpha'}$---and 
because the existence of a stable ground state for the tachyon is 
difficult to see from the perturbative $S$-matrix. While many techniques 
have been employed to tackle this problem (for a review see \cite{SenRev}), 
perhaps the most direct and complete approach uses the formalism of open 
string field theory. For many years, open string field theory provided mostly
a numerical understanding of the tachyon ground state (the ``tachyon vacuum'')
in the level truncation scheme 
\cite{SZ,MT,GR,BSZ,Raey,Iqbal,DeSmet,RaeyThesis}. Then, in 2005 
Schnabl \cite{Schnabl} found an exact solution for the tachyon ground state 
in Witten's open bosonic string field theory \cite{Witten}, providing 
exact formulae for the infinite set of scalar expectation values that arise 
upon tachyon condensation. With these results it was possible to prove 
that the missing energy at the tachyon vacuum exactly corresponds to the 
tension of the unstable D-brane, and that the vacuum supports no 
open string excitations \cite{coh}, precisely as conjectured by Sen 
\cite{Sen1,Sen2,Sen3}.
  
Since then, there has been considerable interest in extending Schnabl's 
results to the superstring. This is not just a matter of principle, but the 
physics of tachyon condensation is much more interesting for the superstring,
revealing a rich spectrum of stable BPS and non-BPS ground states as 
solitons of the string field. Progress on this front however requires 
an analytic solution for the tachyon vacuum in Berkovits' nonpolynomial open 
superstring field theory \cite{BerkovitsI,BerkovitsII}.\footnote{A 
superstring tachyon vacuum solution was found in \cite{supervac} 
using the modified cubic superstring field theory of \cite{PTY,Russians}. 
This solution will play an important role in our analysis, but, for several 
reasons, was not considered to be a definitive solution to the problem of 
tachyon condensation.} Despite many attempts,\footnote{The first attempt 
at analytic solution for the tachyon vacuum was initiated by Berkovits and 
Schnabl \cite{priv}, followed by proposals by Fuchs and Kroyter 
\cite{FK_super}, the author in collaboration with Schnabl \cite{attempt}, 
and possibly others. These solutions proved either to 
be singular or computationally intractable.} no tractable analytic 
solution has been found. In this paper we would 
like to finally propose such a solution.

The solution is constructed as a product of two factors. Each factor 
belongs to a distinct subgroup of the open string star algebra which 
is isomorphic to the group of invertible $2\times 2$ matrices. The 
first factor has matrix entries belonging to the abelian algebra of 
wedge states \cite{RZwedge}, and the second factor has matrix entries 
belonging to the abelian algebra of wedge states deformed by a nonconformal 
boundary interaction related to the condensation of the zero momentum tachyon. 
Like the (closely related) bosonic
tachyon vacuum of \cite{simple}, the solution requires no explicit 
regularization or phantom term, and the simple algebraic structure 
makes it possible to derive all nonpolynomial expressions 
needed for computation of the action in closed form. Evaluating the 
action recovers the expected tension of a non-BPS D-brane. As an added bonus, 
the solution gives a hint as to how D-brane charges may be encoded in the 
topology of the tachyon vacuum gauge orbit.

This paper is organized as follows. In section \ref{sec:Berkovits} we 
review Berkovits' formulation of open superstring field theory with 
an emphasis on concepts which are important for analytic considerations.
In section \ref{sec:algebra} we introduce and motivate the subalgebra of 
states which we will use to formulate the tachyon vacuum solution. 
In sections \ref{sec:solution} and \ref{sec:EOM}, we introduce the solution,
discuss its basic structure, and prove the equations of motion. In 
section \ref{sec:energy} we compute the nonpolynomial action to derive the 
tension of the non-BPS D-brane, and in section \ref{sec:coefficient} we 
compute the expectation value of the tachyon coefficient at the tachyon 
vacuum. In section \ref{sec:charge}, we argue that stability of the codimension
1 kink on a non-BPS D-brane implies that the tachyon vacuum gauge orbit comes 
in two disconnected pieces, related by a topologically nontrivial gauge 
transformation. We show how the analytic solution provides some
preliminary evidence in favor of this conjecture. We end with some conclusions.

\section{Berkovits' Superstring Field Theory}
\label{sec:Berkovits}

Here we review the basics of Berkovits' open superstring field theory 
\cite{BerkovitsI,BerkovitsII}. The theory uses the RNS formalism to describe 
the off-shell dynamics of an open superstring in the Neveu-Schwartz (NS)
 sector.\footnote{Extensions of the action to the Ramond sector are 
described in \cite{BerkRamond,MichishitaRamond,Brazil}.} 
The string field is 
\begin{equation}\Phi=\mathrm{Lie\ algebra\ element}.\end{equation}
We call this the ``Lie algebra element,'' for reasons which will be clear in 
a moment. The Lie algebra element $\Phi$ is a Grassmann even, ghost
and picture number zero\footnote{We follow the ghost and picture 
number assignment conventions of \cite{BSZ}.} state the NS state space of 
an open superstring quantized in a specifically chosen D-brane background.  
As always in string field theory, the string field $\Phi$ represents 
fluctuations of the D-brane system relative to the chosen background. 
For our calculations, we will work on a non-BPS D-brane of Type II 
superstring theory. The extension to a brane/antibrane pair is 
straightforward. Note that, unlike in the bosonic string, for the superstring
the endpoint of tachyon condensation on a generic unstable brane system 
may not be universal, and needs to be constructed in a case-by-case basis. 
For example, in this paper we do not construct the closed string vacuum on 
a separated brane/antibrane system \cite{Bagchi}. 

In the Berkovits theory it is necessary to bosonize the $(\beta,\gamma)$ 
ghosts of the RNS formalism, following \cite{FMS}:
\begin{equation}\beta(z) = \d\xi e^{-\phi}(z),\ \ \ \ \gamma(z) = \eta 
e^\phi(z).\end{equation}
Importantly, the string field $\Phi$ is in the ``large'' Hilbert space, 
that is, the Hilbert space which includes states proportional to the zero 
mode of the $\xi$ ghost. In particular, this means 
\begin{equation}\eta\Phi \neq 0\ \ \ \ \ (\mathrm{in\ general}),\end{equation}
where $\eta\equiv\eta_0$ is the zero mode of the $\eta$ ghost. Both the 
$\eta$ zero mode and the BRST charge $Q\equiv Q_B$ have trivial cohomology 
in the large Hilbert space as a result of the existence of operators 
satisfying 
\begin{eqnarray}
\eta\cdot\xi(z) \lineup = 1, \\
Q\cdot\Big[ c\,\xi\d\xi\, e^{-2\phi}(z)\Big] \lineup = 1.\label{eq:Qhom}
\end{eqnarray}
Therefore in the large Hilbert space the perturbative spectrum is not given by
the cohomology of $Q$. Rather, the perturbative spectrum comes from solutions 
to the linearized equations of motion,
\begin{equation}\eta Q\Phi = 0,\end{equation}
modulo the linearized gauge invariance,
\begin{equation}\Phi' = \Phi + Q\Lambda + \eta \Pi.\end{equation}
Using this gauge symmetry, we can write physical states in the standard form,
\begin{equation}\Phi\sim \xi c e^{-\phi} \mathcal{O}^{\mathrm{m}}(0),
\label{eq:onshell}\end{equation}
where $\mathcal{O}^\mathrm{m}(z)$ is a superconformal matter primary of 
dimension $1/2$. Note that if we drop the $\xi$, this is the same as  
the vertex operator for an on-shell state in the $-1$ picture.

Berkovits' string field theory is constructed using 
Witten's associative star product and open string integration \cite{Witten}. 
Usually we will write the star product without the star 
$AB\equiv A*B$, and we write Witten's integration as a trace $\Tr[\cdot]$ to 
avoid confusion with other integrations. The product, the trace, and the 
differentials $\eta$ and $Q$, satisfy the usual ``axioms'':
\begin{eqnarray}
\mathrm{Nilpotency:}\ \ \ \ \ \  \lineup Q^2 =\eta^2=[Q,\eta]=0;\nonumber\\
\mathrm{Derivation:}\ \ \ \ \ \ \lineup Q(AB) 
= (QA)B+(-1)^A A(QB),\nonumber\\ \lineup
\eta(AB) = (\eta A)B+(-1)^A A(\eta B);\nonumber\\
\mathrm{Integration\ by\ parts:}\ \ \ \ \ \ \lineup \Tr[QA] = 
\Tr[\eta A]=0;\nonumber\\
\mathrm{Associativity:} \ \ \ \ \ \ \lineup A(BC)
 =(AB)C;\nonumber\\
\mathrm{Cyclicity:}\ \ \ \ \ \ \lineup 
\Tr[AB] =(-1)^{AB}\Tr[BA];\label{eq:axioms}
\end{eqnarray}
where $A,B$ and $C$ are generic NS string fields. Though it is not strictly
necessary, we will also freely assume the existence of a star algebra identity
(the identity string field) $1\equiv|I\rangle$ satisfying 
\begin{equation}1*A=A*1=A.\end{equation}
Since all nonzero correlators reduce to
\begin{equation}\left\langle \xi(z)c\d c \d^2c(w)e^{-2\phi}(y)\right\rangle
\equiv 2,\label{eq:basic}\end{equation}
the trace $\Tr[\,\cdot\,]$ is only nonvanishing on states with ghost number 
two and picture number minus one.\footnote{In general \eq{basic} should 
be multiplied by a normalization for the matter correlator. For our purposes 
it is convenient to set this normalization to one. We also set $\alpha'=1$.}

Berkovits' string field theory is defined by a Wess-Zumino-Witten-like 
action \cite{WZW} for the Lie algebra element $\Phi$. To write the action, 
it is helpful to define a ``group element'' $g$ by exponentiating $\Phi$:
\begin{equation}g=e^\Phi=\mathrm{group\ element}.\end{equation}
The group element has a star algebra inverse, 
\begin{equation}g^{-1} = e^{-\Phi}.\end{equation}
Usually the group element $g$ is a more natural field variable for analytic 
calculations. However, unlike $\Phi$, the group element $g$ is constrained 
by the requirement that it must have an inverse. To write the WZW-like action, 
we need to introduce an (arbitrary) continuous 1-parameter family of 
group elements $g(t),\ t\in[0,1]$ interpolating between the identity string 
field $1$ and the dynamical field $g$:
\begin{equation} g(0)=1;\ \ \ g(1)=g=e^\Phi.
\label{eq:bdry}\end{equation}
Next we define three ``connections''  
\begin{equation}\Psi_Q \equiv g(t)^{-1}Qg(t),\ \ \ \ \ \  
\Psi_\eta \equiv g(t)^{-1}\eta g(t),\ \ \ \ \ \ 
\Psi_t \equiv g(t)^{-1}\d_t g(t).
\end{equation}
Then the (standard) WZW-like action takes the form:\footnote{We set the 
open string coupling constant to $1$.} 
\begin{equation}S = -\frac{1}{2}\int_0^1 dt\,\Tr\Big(
\d_t(\Psi_\eta\Psi_Q)
+\Psi_t[\Psi_\eta,\Psi_Q]\Big),\label{eq:action1}
\end{equation}
where $[A,B]\equiv AB -(-1)^{AB}BA$ is the graded commutator. We will 
find it useful to work with a different form of the action, introduced by 
Berkovits, Okawa, and Zwiebach \cite{heterotic}:
\begin{equation}S=-\int_0^1 dt\Tr[(\eta\Psi_t)\Psi_Q].\label{eq:BOZ}
\end{equation}
Though it is not manifest, the action is independent
of the choice of interpolation $g(t)$ provided the boundary 
conditions \eq{bdry} at $t=0$ and $t=1$ are held 
fixed.\footnote{To be precise, the action is {\it locally} independent 
of the choice of interpolation $g(t)$ between $1$ and $g$. We will ignore 
the possibility that there might be distinct homotopy classes of 
interpolations.} Therefore, the action 
only depends on $g$, or equivalently, the Lie algebra element $\Phi$. 
The action is invariant under infinitesimal gauge transformations,
\begin{equation}g' = g+vg+gu,\label{eq:gauge_inv}\end{equation}
where $Qv=0$ and $\eta u=0$. The finite gauge transformation takes the form
\begin{equation}g'=VgU,\end{equation}
where $V$ and $U$ are $Q$- and $\eta$-closed group elements, 
respectively. Finally, the stationary points of the action satisfy 
\begin{equation}\eta(g^{-1}Qg)=0.\label{eq:EOM}\end{equation}
These are the classical equations of motion. 

It can be tricky to prove gauge invariance and derive the equations of 
motion from the action. Let's briefly explain how this is done, following 
closely \cite{heterotic}. Given any derivation $D$ we can define a connection,
\begin{equation}\Psi_D \equiv g(t)^{-1}Dg(t).\end{equation}
By construction this is a flat connection, so for any pair of (anti)commuting 
derivations $D_1$ and $D_2$, the associated field strength must vanish. This 
implies 
\begin{equation}D_1\Psi_{D_2} = (-1)^{D_1D_2}D_2'\Psi_{D_1},\label{eq:vFT}
\end{equation}
where the prime denotes the covariant derivative:
\begin{equation}D'\equiv D+[\Psi_D,\cdot].\end{equation}
Let's denote the variational derivative by $\delta$. With a little algebra one 
can prove the identity:
\begin{equation}\delta\{\eta\Psi_t,\Psi_Q\}-\eta\{\delta\Psi_t,\Psi_Q\}
=-\d_t'\{Q'\Psi_\delta,\Psi_\eta\} +Q'\{\d_t'\Psi_\delta,\Psi_\eta\},
\label{eq:var_id}\end{equation}
where $\{A,B\}\equiv AB+(-1)^{AB}BA$ is the graded 
anticommutator.\footnote{This is a special case of the identity 
\begin{equation}R(D_1,D_2,D_3,D_4) = 
(-1)^{(D_1+D_2)(D_3+D_4)}R(D_3',D_4',D_1,D_2),\end{equation}
where
\begin{equation}R(D_1,D_2,D_3,D_4)\equiv D_1\{D_2\Psi_{D_3},\Psi_{D_4}\}
-(-1)^{D_1D_2}D_2\{D_1\Psi_{D_3},\Psi_{D_4}\}. \end{equation}
Note also that $R$ is graded antisymmetric upon interchange of the first two 
or last two entries.} 
Evaluating the trace and integrating $t$ from $0$ to $1$, 
the left hand side of \eq{var_id} gives (twice) the variation of the action 
\eq{BOZ}. The right hand side gives, after integrating the total $t$ 
derivative,
\begin{equation}\delta S = \left.\Tr\big[\Psi_\delta (\eta\Psi_Q)\big]
\right|_{t=1}.\label{eq:delS}\end{equation}
In this way we see that the action depends only on the value of $g(t)$ at 
$t=1$. The trace with $\Psi_\delta$ is nondegenerate, so setting the variation
of $S$ to zero implies the equations of motion $\eta\Psi_Q|_{t=1}=0$. 
Under the infinitesimal gauge transformation \eq{gauge_inv}, 
$\Psi_\delta$ changes as  
\begin{equation}\Psi_\delta|_{t=1} = \eta(\mathrm{something})
+Q'(\mathrm{something}).\end{equation}
Integration by parts and nilpotency of $\eta$ and $Q'$ then demonstrates gauge
invariance of the action. 

\begin{figure}
\begin{center}
\resizebox{3.7in}{2in}{\includegraphics{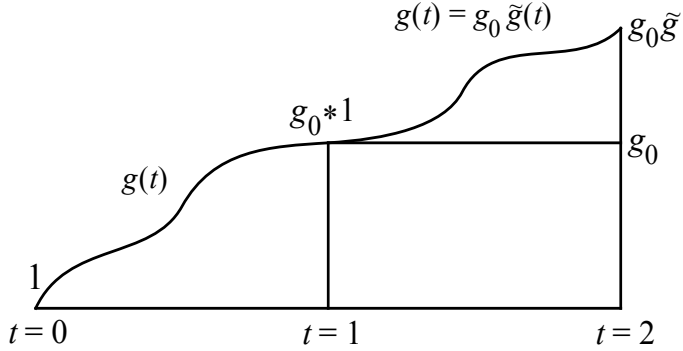}}
\end{center}
\caption{\label{fig:BerkVac4} Choice of interpolating group elements 
$g(t),t\in[0,2]$ for expanding the action around a background solution 
$g_0$.}\end{figure}

Let's explain how to expand the Berkovits theory around a classical solution. 
We write the group element as the product of two factors:
\begin{equation}g = g_0\tilde{g}.\end{equation}
Here the factor $g_0$ is a classical solution which shifts 
from the perturbative vacuum to our new reference background, and 
the factor $\tilde{g}$ describes fluctuations of the field relative to the 
background set by $g_0$. To plug this into the action we must choose a family
of group elements $g(t)$ which interpolates from $1$ to $g_0\tilde{g}$. 
With a reparameterization we can expand the range of $t$ from $0$ to $2$, 
and then choose an interpolation satisfying the conditions
\begin{equation}g(0)=1;\ \ \ g(1)=g_0;\ \ \ g(2)=g=g_0\tilde{g}.\end{equation}
See figure \ref{fig:BerkVac4}. Plugging this in to \eq{BOZ} gives
\begin{equation}S[g]=S[g_0]-\int_1^2 dt\,\Tr[(\eta\Psi_t)\Psi_Q].\end{equation}
The $t\in[0,1]$ region of integration gives the action evaluated on the 
reference solution $g_0$. For $t\in[1,2]$ we can further simplify by writing
$g(t)$ in the form:
\begin{equation}g(t) = g_0\tilde{g}(t),\ \ \ \ t\in[1,2],\end{equation}
where $\tilde{g}(t)$ interpolates from the identity $1$ to the fluctuation
$\tilde{g}$. The $t$-connection $\Psi_t$ evaluated on $g(t)$ is the same as 
that on $\tilde{g}(t)$, since the constant factor $g_0$ cancels out:
\begin{equation}
\Psi_t[g(t)] = \Psi_t[\tilde{g}(t)].
\end{equation}
The $Q$-connection is evaluated as
\begin{equation}
\Psi_Q[g(t)]= \tilde{g}(t)^{-1}(Q+\Psi_0)\tilde{g}(t)
=\Psi_{Q_{\Psi_0}}[\tilde{g}(t)]+\Psi_0. \label{eq:PsiQexp}
\end{equation}
Here 
\begin{equation}\Psi_0\equiv g_0^{-1}Qg_0\label{eq:cubBerk}\end{equation} 
is a solution to the Chern-Simons-like equations of motion of cubic 
superstring field theory \cite{PTY,Russians}
\begin{equation}Q\Psi_0+\Psi_0^2=0,\label{eq:cubEOM}\end{equation}
and 
\begin{equation}Q_{\Psi_0}\equiv Q+[\Psi_0,\cdot]\label{eq:shiftkin}
\end{equation}
is the kinetic operator expanded around the cubic solution $\Psi_0$. 
Since $g_0$ is a solution, $\Psi_0$ is in the small Hilbert space and the 
second term in \eq{PsiQexp} drops out when we plug into the action. 
Now (implicitly) understanding that the connections are evaluated on 
$\tilde{g}(t)$ rather than $g(t)$, and shifting the range of $t$ back to 
$t\in[0,1]$, the action becomes
\begin{equation}S=S[g_0]-\int_0^1 dt\,\Tr[(\eta\Psi_t)\Psi_{Q_{\Psi_0}}].
\end{equation}
Thus the effect of expanding around $g_0$ is to add a constant to the 
action and to replace $Q$ with the kinetic operator $Q_{\Psi_0}$ around
the shifted background.

For states in the GSO($-$) sector the above discussion requires minor 
clarification. The problem is that the zero 
momentum tachyon vertex operator $\xi ce^{-\phi}$ is Grassmann odd, while 
the string field $\Phi$ must be Grassmann even to ensure gauge invariance. 
We solve this problem this by multiplying
the string field with the appropriate Pauli matrix---an ``internal'' 
Chan-Paton factor---determined by the vertex 
operator's Grassmann parity $\eps$ and its worldsheet spinor number $F$ 
\cite{BSZ}:
\begin{center}
\begin{tabular}{|c|c|c|}
\hline
$\ \ \ \eps$\ \ \ &\ \ \ $F$\ \ \ & CP factor
\\
\hline
$0$ & $0$ & $\mathbb{I}$
\\
\hline 
$1$ & $0$ & $\sigma_3$
\\
\hline
$0$ & $1$ & $\sigma_2$
\\
\hline
$1$ & $1$ & $\sigma_1$
\\
\hline
\end{tabular}
\end{center}
For consistency, we also require that all operators acting on the string 
field carry their own internal CP factor according to this table. This means 
that $Q$ and $\eta$ must be multiplied by $\sigma_3$, though in the following 
we will not write this CP factor explicitly, absorbing it into the 
definition of $Q$ and $\eta$. We also assume that the string field trace 
$\Tr[\cdot]$ includes an implicit factor of $1/2$ times a trace 
over internal CP factors. To see how this prescription solves the problem with
GSO($-$) states, it is useful to introduce the concept of 
{\it effective Grassmann parity} 
\cite{exotic}:
\begin{equation}E\equiv \eps+F\ \ \ \ \ (\mathrm{mod}\ 2).\label{eq:eff}
\end{equation}
Effective Grassmann parity helps keep track of signs when commuting
string fields and their associated CP factors past each other. In fact, the 
requisite sign can be described by a ``double graded'' commutator, 
\begin{equation}\llbracket A,B\rrbracket \equiv AB 
-(-1)^{E(A)E(B)+F(A)F(B)}BA.\label{eq:dbracket}\end{equation}
This means that the algebra of string fields on a non-BPS D-brane is like 
an algebra of matrices whose entries contain two {\it mutually commuting} 
types of Grassmann number. The first type has Grassmannality measured by $E$ 
and the second type by $F$. Effective Grassmann parity enters 
the string field theory axioms \eq{axioms} through the relations
\begin{eqnarray}\eta(AB)\lineup = 
(\eta A)B+(-1)^{E(A)}A (\eta B);\\
Q(AB)\lineup = (QA)B+(-1)^{E(A)}A (QB); \\
\Tr[AB]\lineup = (-1)^{E(A)E(B)}\Tr[BA].
\end{eqnarray}
Note that permuting the trace does not produce a sign from the 
parity of $F$, contrary to what we might expect from \eq{dbracket}. 
This is because the half integer conformal dimension of vertex operators 
in the GSO($-$) sector produces an anomalous sign when permuting 
the conformal maps defining the Witten vertex \cite{BSZ}. For this reason, 
worldsheet spinor number plays no role in establishing gauge 
invariance of the action. This means we can incorporate GSO($-$) states by 
assuming that the string field is {\it effective} Grassmann even
\begin{eqnarray}g 
= e^{\Phi} \lineup = \mathrm{\ Effective\ Grassmann\ even}.
\end{eqnarray}
The zero momentum tachyon $\xi c e^{-\phi}$ is ``effectively'' Grassmann 
even, even though it's a Grassmann odd operator. Since worldsheet spinor 
number $F$ does not appear in the string field theory axioms, the 
WZW-like action uses a commutator which is graded only with respect to 
effective Grassmann parity
\begin{equation}[A,B] \equiv AB -(-1)^{E(A)E(B)}BA.
\end{equation} 
This is the commutator which appears in the Wess-Zumino term of 
\eq{action1}, and in the shifted kinetic operator when expanding the action 
around a nontrivial solution \eq{shiftkin}.

\section{Algebra}
\label{sec:algebra}

\begin{figure}
\begin{center}
\resizebox{4.75in}{2.1in}{\includegraphics{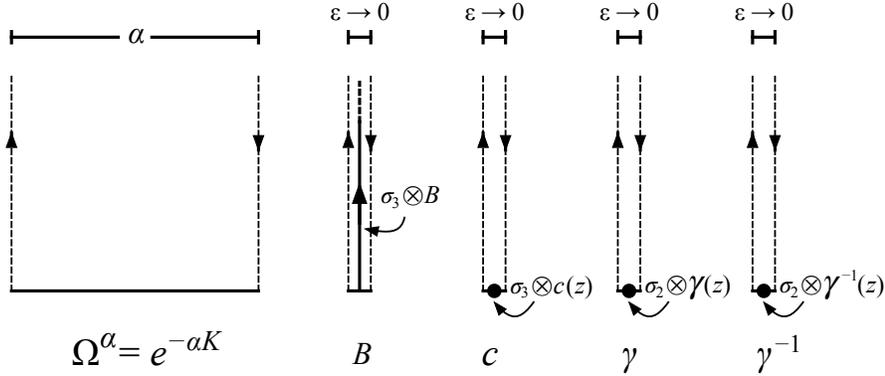}}
\end{center}
\caption{\label{fig:BerkVac5} The wedge state $\Omega^{\alpha}$ and the 
remaining four fields in \eq{alg} as semi-infinite strips 
with operator insertions. The arrows on the vertical edges indicate the 
direction of the parameterization of the open string $\sigma\in[0,\pi]$.}
\end{figure}

The tachyon vacuum is constructed by taking star products of five string 
fields:\footnote{We learned the notation $\gamma^{-1}$ from N. Berkovits and 
M. Schnabl.}
\begin{eqnarray}
K\,\lineup\ \ \ \rightarrow\ \ \ \mathbb{I}\otimes\,K =\mathbb{I}\otimes\int_{-i\infty}^{i\infty}
\frac{dz}{2\pi i}T(z),\nonumber\\
B\,\lineup\ \ \ \rightarrow\ \ \ \sigma_3\otimes B = 
\sigma_3\otimes\int_{-i\infty}^{i\infty}
\frac{dz}{2\pi i}b(z),\nonumber\\
c\ \lineup\ \ \ \rightarrow\ \ \ \sigma_3\otimes c(z),\phantom{\Big(}
\nonumber\\
\gamma\ \lineup\ \ \ \rightarrow\ \ \ \sigma_2\otimes \gamma(z)
=\sigma_{2}\otimes\eta e^\phi(z),\phantom{\Big(}\nonumber\\
\gamma^{-1}\!\lineup\ \ \ \rightarrow\ \ \ \sigma_2\otimes\gamma^{-1}(z)
=\sigma_{2}\otimes e^{-\phi}\xi(z).\phantom{\Big(}\label{eq:alg}
\end{eqnarray}
Here we use the algebraic formalism of Okawa \cite{Okawa}, where 
the string fields $K,B,c,\gamma$ and $\gamma^{-1}$ represent 
corresponding operator insertions (with internal CP factors) in correlation 
functions on the cylinder. To review, we can visualize the definition of 
$K,B,c,\gamma,\gamma^{-1}$ using the Schr\"odinger representation, as 
functionals defined by a worldsheet path integral on a semi-infinite 
vertical ``strip'' with boundary conditions on its vertical edges 
corresponding to the left and right halves of the open string. Specifically, 
$K,B,c,\gamma$ and $\gamma^{-1}$ are defined by a path integral on an 
infinitesimally thin strip containing the appropriate operator insertion, 
as shown in figure \ref{fig:BerkVac5}.\footnote{Explicit Fock space 
expansions of $K,B,c$, and by extension $\gamma,\gamma^{-1}$, can be found 
from other equivalent definitions, for example those provided in 
\cite{SSF1,OkawaRev,Aldo}.} Star multiplication glues the right 
boundary of the first strip to the left boundary of the second 
strip,\footnote{We use the left handed star product convention \cite{simple}.
Operator insertions inside the correlator and internal CP factors are 
multiplied in the order of star multiplication. See appendix A of 
\cite{exotic} for a description of various signs related to the GSO($-$) 
sector in the left handed convention.} and the trace glues the left and right 
boundaries to form a correlation function on the cylinder. If we assume the 
sliver coordinate frame \cite{RZ},\footnote{The choice of coordinate frame 
corresponds to the choice of parameterization of the string on the left and 
right boundaries of the strip \cite{RZO,SSF1}. This choice is only relevant 
for our computation of the tachyon coefficient in section 
\ref{sec:coefficient}.} the field $K$ generates the algebra of wedge states 
\cite{RZwedge,SchWedge}, in that any star algebra power of the 
$SL(2,\mathbb{R})$ vacuum $\Omega\equiv|0\rangle$ can be 
written
\begin{equation}\Omega^\alpha = e^{-\alpha K}, \ \ \ \ \ \alpha\geq 0.
\end{equation}
A wedge state $\Omega^\alpha$ represents a semi-infinite strip of width 
$\alpha$, as shown in figure \ref{fig:BerkVac5}. 
The fields $K,B,c,\gamma,\gamma^{-1}$ come with a list of quantum 
numbers summarized in table \ref{tab:fields}.

Let's explain why this algebraic setup is relevant for the problem of tachyon
condensation. The fields $K,B$ and $c$ in \eq{alg} appear in Schnabl's 
solution \cite{Schnabl} and related solutions \cite{Okawa,SSF1,SSF2,simple} 
for the tachyon vacuum in the bosonic string. The superstring tachyon vacuum
is related to these solutions though equation \eq{cubBerk}, and for this 
reason we will need $K,B$ and $c$ as well. However, we also need two 
additional string fields, $\gamma$ and $\gamma^{-1}$. They live in the 
GSO($-$) sector, and are required to give an expectation value to the 
tachyon on the non-BPS D-brane. In particular, the zero momentum tachyon 
can be written
\begin{equation}
\gamma^{-1}c \,\sim\, 
\xi ce^{-\phi}(0).\end{equation}
Commuting this with $B$ gives $\gamma^{-1}$. The BRST variation 
of $c$ generates $\gamma^2$, so once we have $\gamma^{-1}$ we need $\gamma$ 
as well.  Therefore the fields $K,B,c,\gamma$ and $\gamma^{-1}$ give the 
minimum extension of the bosonic framework sufficient to describe tachyon 
condensation on a non-BPS D-brane in superstring theory.

\begin{table}
\renewcommand{\arraystretch}{1.0}
\begin{center}
\begin{tabular}{|c|c|c|c|c|c|c|c|c|}
\hline 
\phantom{\Bigg(}
field\ \ \ 
& \!\!${\mathrm{ghost} \atop \mathrm{number}}$\!\!
& \!\!${\mathrm{picture} \atop \mathrm{number}}$\!\!
& \!\!${\mathrm{effective}\atop \mathrm{Grassmann\ parity}}$\!\!
& \!\!${\mathrm{worldsheet}\atop \mathrm{spinor\ number}}$\!\!
& \!\!${\mathrm{scaling}\atop \mathrm{dimension}}$\!\!
& reality
& \!\!Twist\!\!
& ${\mathrm{Hilbert}\atop\mathrm{space?}}$
\\
\hline \!\!\!\!\!\phantom{\Bigg(} $K$ 
& $0$
& $0$
& even
& $0$
& $1$
& real
& $1$
& small
\\
\hline  \!\!\!\!\!\phantom{\Bigg(} $B$ 
& $-1$
& $0$
& odd
& $0$
& $1$
& real
& $1$
& small
\\ 
\hline \!\!\!\!\!\phantom{\Bigg(}  $c$ 
& $1$
& $0$
& odd
& $0$
& $-1$
& real
& $-1$
& small
\\  
\hline \!\!\!\!\!\phantom{\Bigg(} $\gamma$ 
& $1$
& $0$
& odd
& $1$
& $-\frac{1}{2}$
& real
& $-i$
& small
\\ 
\hline \!\!\!\!\!\phantom{\Bigg(} $\gamma^{-1}$ 
& $-1$
& $0$
& odd
& $1$
& $\frac{1}{2}$
& real
& $i$
& large 
\\
\hline  \!\!\!\!\!\phantom{\Bigg(} $\zeta$ 
& $0$
& $0$
& even
& $1$
& $-\frac{1}{2}$
& \!\! imaginary\!\!
& $-i$
& large
\\ 
\hline \!\!\!\!\!\phantom{\Bigg(} $V$ 
& $0$
& $0$
& even
& $1$
& $\half$
& real
& $-i$
& large
\\ 
\hline
\end{tabular}
\end{center}
\caption{\label{tab:fields} Table of useful quantum numbers for fields 
\eq{alg} and \eq{comp}. ``Scaling dimension'' refers to the eigenvalue 
under the scaling generator in the sliver frame $\half\mathcal{L}^-$ 
\cite{IdSing}; ``Reality'' refers to the eigenvalue under reality 
conjugation \cite{tensor,exotic} (with real or imaginary meaning 
$A^\ddag = \pm A$); Twist refers to the eigenvalue under twist conjugation 
$A^\S \equiv e^{i\pi L_0}A$ \cite{simple,exotic}.} \end{table}

The fields \eq{alg} satisfy a number of algebraic relations:
\begin{eqnarray}\lineup B^2=c^2=0,\ \ \ \ \ \ \ 
\gamma\gamma^{-1}=\gamma^{-1}\gamma= 1;\nonumber\\
\lineup [K,B] = 0,\ \ \ \ \ [B,c]=1,\ \ \ \ \ [B,\gamma]=0,\ \ \ \ \ 
[B,\gamma^{-1}]=0;
\nonumber\\
\lineup [K,(\mathrm{anything})] = \d(\mathrm{anything});\nonumber\\
\lineup \llbracket \,(c \mathrm{\ and/or\ gamma\ ghosts})\,,\,
(c \mathrm{\ and/or\ gamma\ ghosts})\,\rrbracket = 0.
\label{eq:algid}\end{eqnarray}
The second to the last equation means that the commutator of $K$ with a 
string field computes the worldsheet derivative of its corresponding 
operator insertion in correlation functions on the cylinder. The last equation 
means that any two string fields made from products of $c,\gamma,\gamma^{-1}$ 
and worldsheet derivatives thereof always commute in the sense of the double 
bracket \eq{dbracket}.\footnote{The last equation seems analogous to the 
``auxiliary identities'' in the $K,B,c$ subalgebra of the bosonic string 
\cite{IdSing}. However, this analogy is imprecise since there are no 
automorphisms of the $K,B,c,\gamma,\gamma^{-1}$ subalgebra like those in the 
$K,B,c$ subalgebra \cite{simple_talk,IdSing} which preserve  ``fundamental'' 
but not ``auxiliary'' algebraic relations.  Such automorphisms however can be 
defined on a larger subalgebra generated by products of 
$G,B,c,\gamma^2,\alpha=-c\gamma^{-2}$, considered in \cite{exotic}. As
 discussed in 
\cite{MNT,HataKojita,Masuda,Aldo2,HataKojita2}, these automorphisms may have 
applications in the search for multiple brane solutions \cite{MS,MS2}.} 
We have BRST variations:
\begin{eqnarray}\lineup QK= 0,\nonumber\\
\lineup QB=K,\nonumber\\
\lineup Qc = c\d c-\gamma^2,\nonumber\\
\lineup Q\gamma = c\d \gamma -\frac{1}{2}\d c \gamma,\nonumber\\
\lineup Q\gamma^{-1} = c\d \gamma^{-1} +\frac{1}{2}\d c \gamma^{-1}.
\label{eq:algBRST}\end{eqnarray}
Note the order of multiplication of $c$ and $\gamma,\gamma^{-1}$ 
matters in these equations, since $\gamma$ and $\gamma^{-1}$ are 
effective Grassmann odd (despite the fact that the $\gamma$-ghost is 
bosonic). All fields are annihilated by the eta zero mode except for 
$\gamma^{-1}$:
\begin{equation}\eta\gamma^{-1}\neq 0.\end{equation}
The field $\eta\gamma^{-1}$ is outside the $K,B,c,\gamma,\gamma^{-1}$ 
subalgebra. Note that $\eta\gamma^{-1}$ has singular OPE with 
$\gamma^{-1}$, so we must be careful how it appears in star 
products with other states. 

It is useful to introduce two composite string fields:
\begin{eqnarray}
\zeta \equiv \gamma^{-1}c\ \ \ \ \lineup\ \ \ \rightarrow\ \ \ 
i\sigma_1\otimes \zeta(z)\equiv i\sigma_1\otimes \gamma^{-1}c(z),\\
V \equiv\frac{1}{2}\gamma^{-1}\d c \lineup\ \ \ \rightarrow\ \ \ 
i\sigma_1\otimes V(z)\equiv i\sigma_1\otimes\frac{1}{2}\gamma^{-1}\d c(z).
\label{eq:comp}\end{eqnarray}
The first field $\zeta$ is the zero momentum tachyon. The second field $V$ 
can be interpreted as a kind of ``integrated vertex operator'' associated 
with the zero momentum tachyon. To see why, consider the relation 
\begin{equation}Q\zeta = cV +\gamma.\end{equation}
If $\zeta$ were an on-shell state of the form \eq{onshell}, then the operator 
multiplying $c$ above would be the integrated vertex operator which 
generates a boundary deformation of the worldsheet action associated with 
this on-shell state. Of course $\zeta$ is off-shell, but $V$ can still be 
viewed as an ``integrated vertex operator'' generating a 
nonconformal boundary deformation on the worldsheet. In a moment we will see 
how this interpretation is borne out in the solution. The fields satisfy 
some useful identities:
\begin{eqnarray}
\lineup [B,Q\zeta] =V,\nonumber\\ 
\lineup [B,Qc]=\d c,\nonumber\\
\lineup \zeta^2 = 0\nonumber\\
\lineup c\zeta = \zeta c =0,\nonumber\\
\lineup \gamma \zeta = -\zeta\gamma = c,\nonumber\\
\lineup (Q\zeta)\zeta = -\zeta(Q\zeta) = c,\nonumber\\
\lineup (Qc)\zeta = \zeta (Qc) = -\gamma c,\nonumber\\
\lineup (Q\zeta)c = -c(Q\zeta) = \gamma c,\nonumber\\
\lineup (Q\zeta)^2 = -Qc. \label{eq:id2}
\end{eqnarray}
These identities follow immediately from \eq{comp}, \eq{algid} 
and \eq{algBRST}, but appear often enough in computations to be worth 
remembering.

\section{Solution}
\label{sec:solution}

\begin{figure}
\begin{center}
\resizebox{3.8in}{3.8in}{\includegraphics{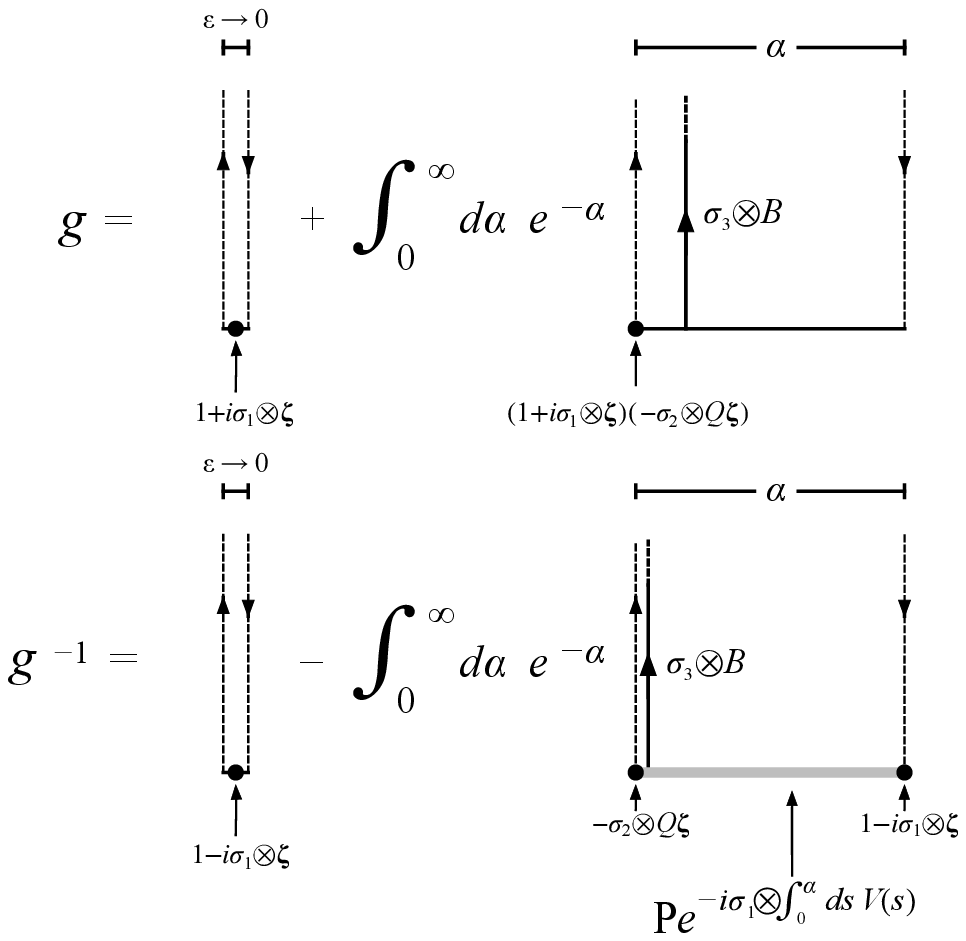}}
\end{center}
\caption{\label{fig:BerkVac8} Worldsheet picture of the solution \eq{BerkVac} 
and \eq{BerkVacinv} as strips appearing inside correlation functions on 
the cylinder.}
\end{figure}

The tachyon vacuum solution takes the form
\begin{equation}g = (1+\zeta)\left(1+Q\zeta\frac{B}{1+K}\right),
\label{eq:BerkVac}\end{equation}
or, in terms of the inverse group element,
\begin{equation}g^{-1} = \left(1-Q\zeta\frac{B}{1+K+V}\right)(1-\zeta).
\label{eq:BerkVacinv}
\end{equation}
See figure \ref{fig:BerkVac8} for a worldsheet picture of the solution. 
A Berkovits solution always defines a corresponding solution to the 
Chern-Simons equations of motion \eq{cubEOM}, and in our case the solution is:
\begin{equation}\Psi=g^{-1}Qg = c-Qc\frac{B}{1+K}.
\label{eq:simple}\end{equation}
This is the ``simple'' analytic solution for tachyon condensation found
in \cite{simple}.\footnote{The solution of \cite{simple} was proposed in 
the context of Witten's open bosonic string field theory, but it translates
to the Chern-Simons superstring essentially unchanged 
\cite{supervac,Gorb,Aldo}.} Since $\Psi$ is in the small Hilbert space,
\eq{simple} implies that $g$ satisfies the Berkovits equations of 
motion. We will give a more detailed demonstration in section \ref{sec:EOM}. 
Note that we can automatically obtain another tachyon vacuum solution by 
making a parity flip $(-1)^F$ in the GSO($-$) sector. This is the solution 
on the ``other side'' of the perturbative vacuum in the tachyon effective 
potential.

A characteristic feature of this solution is the presence of a peculiar 
nonconformal boundary interaction generated by $V$, as can be seen in the 
expression for $g^{-1}$ in equation \eq{BerkVacinv}. Let's explain this in 
more detail. The factor $\frac{1}{1+K}$ can be defined using
the Schwinger parameterization as an integral over wedge states:
\begin{equation}\frac{1}{1+K} = \int_0^\infty d\alpha\, e^{-\alpha}\,
\Omega^\alpha.\end{equation}
Likewise, the factor $\frac{1}{1+K+V}$ which appears in the inverse group 
element \eq{BerkVacinv} can be defined as an integral over ``deformed'' 
wedge states: 
\begin{equation}\frac{1}{1+K+V} = \int_0^\infty d\alpha\, 
e^{-\alpha} e^{-\alpha(K+V)}.\end{equation}
The ``deformed'' wedge state $e^{-\alpha(K+V)}$ corresponds to a strip of 
width $\alpha$ carrying an infinite number of boundary insertions of $V$. 
As shown in \cite{KOS}, the insertions arrange themselves in such a way as
to add a boundary coupling to the worldsheet action. In our case the boundary
coupling inserts a nonlocal exponential insertion of the form:
\begin{equation}e^{-\alpha(K+V)}\ \ \ \rightarrow\ \ \ 
\mathcal{P}\exp\left[-i\sigma_1\otimes\frac{1}{2}\int_0^\alpha ds\,
\gamma^{-1}\d c(s)\right]. 
\label{eq:Vins}\end{equation}
where we assume that the strip of width $\alpha$ has its right vertical 
edge aligned with the imaginary axis $\mathrm{Re}(z)=0$. The insertion in 
\eq{Vins} is path ordered since $\gamma^{-1}\d c$ is a fermionic operator. 
We define the ordering in sequence of decreasing position on the real axis
\begin{eqnarray}\mathcal{P}\gamma^{-1}\d c(s_1)\gamma^{-1}\d c(s_2)...
\gamma^{-1}\d c(s_n) \lineup 
\equiv \gamma^{-1}\d c(s_{i_1})\gamma^{-1}\d c(s_{i_2})...
\gamma^{-1}\d c(s_{i_n}),\nonumber\\
\lineup \ \ \ \ \ \ \ \ \ \ \ \ \ \ \ \ \ \ 
\ \ \ \ \ \ (s_{i_1}>s_{i_2}>...>s_{i_n}).
\end{eqnarray}
An interesting property of this boundary interaction is that it is BRST 
invariant. In particular, we have the property
\begin{equation}Qe^{-\alpha(K+\lambda V)} =-\lambda( Q\zeta +\lambda c)
e^{-\alpha(K+\lambda V)}+
e^{-\alpha(K+\lambda V)}\lambda(Q\zeta +\lambda c),
\end{equation} 
where $\lambda$ is the coupling constant of the deformation. This means that
the only contribution to the BRST variation occurs at the interface between
the deformed and undeformed boundary condition. 
This might suggest that we could define the boundary interaction
in terms of boundary condition changing operators \cite{KOS}, but this language
is not quite appropriate since the boundary interaction is nonconformal. 
Note that the conservation of $bc$ ghost number implies that the
boundary interaction contributes at most a finite number of insertions to 
any particular correlator. This means that \eq{BerkVacinv} 
is a manifestly finite and explicitly computable state in the Fock space 
expansion.

The Berkovits solution has a number of formal similarities with the ``simple'' 
solution \eq{simple} of the bosonic string. One curious similarity
is that both solutions can be written as a linearized gauge transformation
of the zero momentum tachyon expressed in a particular gauge (specifically,
the dressed-Schnabl gauge of \cite{simple}). This 
can be seen from the expressions
\begin{eqnarray}\Psi \lineup = c\frac{1}{1+K} -Q\left(c\frac{B}{1+K}\right),
\label{eq:simple2}\\
g \lineup = 1+\zeta\frac{1}{1+K} 
+Q\left(\zeta\frac{B}{1+K}\right)-c\frac{B}{1+K}.\label{eq:BerkVac2}
\end{eqnarray}
Here $c\frac{1}{1+K}$ is the zero momentum tachyon of the bosonic string,
and the second term in \eq{simple2} is BRST exact. The state 
$1+\zeta\frac{1}{1+K}$ represents a deformation of the perturbative vacuum $1$
by the zero momentum tachyon $\zeta\frac{1}{1+K}$, and the last two terms 
in \eq{BerkVac2} are $Q$- and $\eta$-exact, respectively. Another 
similarity is that
$g$ and $\Psi$ do not need to be defined with a regularization and 
``phantom term'' \cite{phantom}, unlike Schnabl's solution for the 
bosonic string \cite{Schnabl,simple}. While this is an advantage, the down 
side is that these solutions are close to being singular from the 
perspective of the identity string 
field. This observation can be formalized in the dual $\mathcal{L}^-$ level
expansion \cite{IdSing}, where $g$ takes the form
\begin{equation}g=1+Q\left(\zeta\frac{B}{K}\right)+\mathrm{lower\ levels}.
\end{equation}
Taking the logarithm gives the dual $\mathcal{L}^-$ level expansion of 
the Lie algebra element:
\begin{equation}\Phi = Q\left(\zeta\frac{B}{K}\right)+\mathrm{lower\ levels}.
\end{equation}
Consulting table \ref{tab:fields}, we see that the leading level in this 
expansion is $-\frac{1}{2}$. Since the trace of the Lie algebra element can 
be used to define the on-shell part of the boundary state 
\cite{Ellwood,Michishita}, this is the highest half-integer level consistent
with a regular solution \cite{IdSing}. So, in a sense, the Berkovits solution
\eq{BerkVac} is as identity-like as possible given constraints of regularity.

For later analysis it will be useful to consider a slight generalization of 
the solution \eq{BerkVac}. To obtain this generalization, note that since 
$\Psi$ is in the GSO($+$) sector, both the GSO($+$) part and the 
GSO($-$) part of $g$ satisfy
\begin{equation}
Qg_+ = g_+\Psi,\ \ \ \ Qg_- = g_-\Psi.
\end{equation}
This almost implies that $g_+$ and $g_-$ are separately solutions to the 
equations of motion, except (it turns out\footnote{The $g_+$ component is 
Okawa's left gauge transformation from the perturbative vacuum to 
the tachyon vacuum  \cite{Okawa}, and is not invertible (cf. \cite{Integra}). 
The $g_-$ component is more subtle, and is possibly an interesting solution in 
its own right. However, the inverse of $g_-$ is a fairly singular state in 
the dual $\mathcal{L}^-$ level expansion \cite{IdSing}. Earlier collaboration
with M. Schnabl \cite{attempt} investigated a similar solution, 
but problems with the identity string field ultimately rendered it 
unmanageable.}) that $g_+$ and $g_-$ are not invertible. Still, we can try 
to form a solution by taking a linear combination of $g_+$ and $g_-$:
\begin{equation}Q(pg_++qg_-) = (pg_++qg_-)\Psi.\end{equation}
Imposing regularity of $\Phi$ in the dual $\mathcal{L}^-$ expansion requires 
$p=1$. Since $g_+$ is not by itself a solution, we must add some $g_-$ with 
$q\neq 0$. This defines a class of tachyon vacuum solutions generalizing 
\eq{BerkVac}:
\begin{eqnarray}
g\lineup = (1+q\zeta)\left(1-(1-q^2)c\frac{B}{1+K}
+qQ\zeta \frac{B}{1+K}\right),\label{eq:BerkVacq}\\
g^{-1}\lineup = \left(1+(1-q^2)c\frac{B}{q^2+K+qV}-qQ\zeta\frac{B}{q^2+K+qV}
\right)(1-q\zeta).\label{eq:BerkVacqinv}
\end{eqnarray}
All of these solutions describe the tachyon vacuum, and $q$ is merely a 
gauge parameter which roughly corresponds to the expectation value of the 
tachyon. We will clarify this relation in section \ref{sec:coefficient}.

Let's explain why the solution \eq{BerkVac} allows an analytic proof of Sen's
conjecture, whereas other proposals have proven intractable. 
Consider a possible tachyon vacuum solution of the form
\begin{equation}g = 1+\zeta B\frac{K\Omega}{1-\Omega}c\Omega + 
Q\big(\zeta B\Omega\big)-cB\Omega.\label{eq:SchVac}\end{equation}
Replacing $\Omega\to\frac{1}{1+K}$ gives back our solution 
\eq{BerkVac} as expressed in \eq{BerkVac2}. Formally \eq{SchVac} satisfies
\begin{equation}g^{-1}Qg = \left(c\frac{KB}{1-\Omega}c+B\gamma^2\right)\Omega.
\end{equation}
The right hand side is Schnabl's solution \cite{Schnabl}, with superstring 
correction \cite{supervac} and ``security strip'' placed on the right. The 
problem comes with defining $g^{-1}$. Since $g$ can be written as 
$1+(\mathrm{something})$ we can try to define $g^{-1}$ as a geometric series: 
\begin{equation}g^{-1}= 1-(\mathrm{something})+(\mathrm{something})^2-...\ .
\label{eq:Schseries}\end{equation}
At each order the number of cross terms appearing in $(\mathrm{something})^n$
grows exponentially with $n$. Aside from the practical difficulty of actually 
computing this series, the perturbative expansion is not controlled 
by a parametrically small parameter, so it is not clear whether the sum 
meaningfully converges. And without a usable definition of $g^{-1}$, it seems
impossible to evaluate the action and prove Sen's conjecture. This is a 
problem common to many tachyon vacuum solutions in the 
$K,B,c,\gamma,\gamma^{-1}$ subalgebra, and has been a major obstacle to 
analytic solution.

While the solution we have found has many nice properties, there are several 
notable shortcomings:
\begin{itemize}
\item As far as we know, the solution is not defined by a linear
gauge condition. Actually, we are not certain how to implement an acceptable 
gauge fixing in our framework, since the $(\xi_0=0)$-gauge used in level 
truncation studies \cite{BSZ,Raey} does not fit well with the 
$K,B,c,\gamma,\gamma^{-1}$ subalgebra we have been using. This means
in particular that we have no natural definition of the tachyon 
potential, though to be honest even in the bosonic string the Fock space 
tachyon potential in Schnabl gauge has not yet been computed.
\item The solution \eq{BerkVac} fails to satisfy the string field reality 
condition \cite{tensor,Ohmori,supermarg,exotic}\footnote{Determining the 
correct sign in the reality condition on GSO($-$) states is a little tricky 
\cite{Ohmori,exotic}. Assuming $^\ddag$ is defined as the composition of 
Hermitian followed by BPZ conjugation \cite{exotic}, equation \eq{reality} 
is correct in the left-handed star product convention which we have been 
using. In the right handed convention favored by \cite{BSZ,Ohmori,Okawa} 
the correct reality condition is $g^\ddag = (-1)^F g^{-1}$.}
\begin{equation}g^\ddag = g^{-1}.\label{eq:reality}\end{equation}
This is a nonlinear condition on the group element which is typically 
difficult to solve. Formally we can construct a real 
solution as a gauge transformation of \eq{BerkVac} as follows: Take $g$ 
and define a new solution $\tilde{g} \equiv \frac{1}{\sqrt{1+K}}g\sqrt{1+K}$. 
Then define a third solution \cite{supermarg}
\begin{equation}\hat{g}\equiv \frac{1}{\sqrt{\tilde{g}\tilde{g}^\ddag}}
\tilde{g}.
\end{equation}
This solution formally satisfies $\hat{g}^\ddag=\hat{g}^{-1}$ as desired. 
The problem is that this solution is complicated, and we know little about 
it aside from its formal definition. Perhaps other approaches, such as 
\cite{Okawa_real_super_marg}, could be adapted to find a more tractable real 
solution.
\item As mentioned before, the solution is fairly identity-like.  This
suggests that the energy may not be very well behaved in the level expansion 
\cite{simple}. A related problem is that the string field 
\begin{equation}g^{-1}\eta g\end{equation}
is logarithmically divergent due to an integrated collision between 
$V$ and $\eta\zeta$. Fortunately this divergence is 
absent in the computation of observables. 
\end{itemize} 
None of these issues will prove to be fatal for our purposes. But we hope 
that the solution presented here is a starting point for finding other 
solutions which address some of these problems, or possibly have other 
interesting properties. 

\section{Equations of Motion}
\label{sec:EOM}

Now we will prove the equations of motion for the solution \eq{BerkVac}. 
We prove the equations of motion in two steps. First we show that the 
expressions \eq{BerkVac} and \eq{BerkVacinv} are actually inverses of 
one another:
\begin{equation}g^{-1}g = gg^{-1}=1.\label{eq:inverses}\end{equation}
Second we will show that $g$ satisfies
\begin{equation}Qg = g\Psi.\label{eq:naiveEOM}\end{equation}
This together with \eq{inverses} implies \eq{simple}, which implies the 
equations of motion.

First let's prove that $g$ and $g^{-1}$ are inverses by direct 
computation with the identities \eq{id2}:
\begin{eqnarray}g^{-1}g\lineup = \left(1-Q\zeta\frac{B}{1+K+V}\right)(1-\zeta)
(1+\zeta)\left(1+Q\zeta\frac{B}{1+K}\right),\nonumber\\
\lineup = 1-Q\zeta\frac{B}{1+K+V}+Q\zeta\frac{B}{1+K}-
Q\zeta\frac{B}{1+K+V}Q\zeta\frac{B}{1+K},\nonumber\\
\lineup  = 1-Q\zeta\frac{B}{1+K+V}+Q\zeta\frac{B}{1+K}-
Q\zeta\frac{B}{1+K+V}V\frac{1}{1+K}.
\label{eq:gginv1}\end{eqnarray}
Now look at the $V$ stuck in the middle of the last term. Write it as the 
difference of two terms:
\begin{equation}V=(1+K+V)-(1+K).\end{equation}
The terms cancel against one of the two factors on either side of 
$V$ in \eq{gginv1}
\begin{equation}
g^{-1}g = 1-Q\zeta\frac{B}{1+K+V}+Q\zeta\frac{B}{1+K}-Q\zeta
\left(\frac{B}{1+K}-\frac{B}{1+K+V}\right).
\end{equation}
What's left cancels leaving $g^{-1}g=1$.

While this proof is sufficient, there is another way to look at this which 
provides more insight. Consider a class of states
\begin{equation}M = -\gamma BX_1\zeta+cBX_2+\gamma BY_1-cB Y_2 \zeta,
\label{eq:2by2}\end{equation}
where $X_1,X_2$ and $Y_1,Y_2$ are any string fields which commute with $B$. 
It turns out that states of this form multiply like $2\times 2$ matrices 
with $X_1,X_2$ and $Y_1,Y_2$ placed in the entries as follows:\footnote{This 
matrix structure generalizes an old idea of Schnabl \cite{priv} 
for building the tachyon vacuum starting from a ``square root'' of the 
identity string field: $\sqrt{1} =\zeta +B\gamma$. The author thanks him 
for sharing this insight.}
\begin{equation}M = \left(\begin{matrix} X_1 & Y_1 \\ Y_2 & X_2 \end{matrix}
\right).\end{equation}
Multiplying out factors and calculating $Q\zeta$,  we can express 
the solution \eq{BerkVac} in the form
\begin{equation}g = -\gamma B\zeta +cB(K+V)\frac{1}{1+K}
+\gamma B\frac{1}{1+K} +cB\zeta .\end{equation}
Comparing with \eq{2by2} we find that the solution can be expressed as 
a $2\times 2$ matrix:
\begin{equation}g = \left(\begin{matrix}1 & {\displaystyle \frac{1}{1+K}}
 \\ -1 & {\displaystyle (K+V)\frac{1}{1+K}}\end{matrix}\right).\end{equation}
Since the entries of this matrix do not commute, 
computing the inverse could be difficult. Luckily $g$ can be factorized:
\begin{equation}g = \left(\begin{matrix}1 & 1 
 \\ -1 & K+V \end{matrix}\right)\left(\begin{matrix} 1 & 0 \\ 0 & 
{\displaystyle \frac{1}{1+K}}\end{matrix}\right),\label{eq:BerkVacmat}
\end{equation}
and the entries in each individual matrix factor commute. Thus we have 
expressed the solution as the product of two factors in two noncommuting 
copies of the group of invertible $2\times 2$ matrices $GL_2$. Computing
the inverse of $g$ is now as easy as computing the inverse of a $2\times 2$ 
matrix:
\begin{equation}g^{-1} = \left(\begin{matrix} 1 & 0 \\ 0 & 1+K\end{matrix}
\right)\left(\begin{matrix} 1-\frac{1}{1+K+V} & -\frac{1}{1+K+V} \\ 
\frac{1}{1+K+V} & \frac{1}{1+K+V}\end{matrix}\right).
\end{equation}
With a few more steps we can obtain the more familiar expression for $g^{-1}$:
\begin{eqnarray}
g^{-1} \lineup = \left(\begin{matrix} 1-{\displaystyle \frac{1}{1+K+V}} & 
{\displaystyle -\frac{1}{1+K+V}} \\ 
{\displaystyle (1+K)\frac{1}{1+K+V}} & {\displaystyle (1+K)\frac{1}{1+K+V}}
\end{matrix}\right),\nonumber\\
\lineup =\left(\begin{matrix} 1-{\displaystyle \frac{1}{1+K+V}} & 
{\displaystyle -\frac{1}{1+K+V}} \\ 
{\displaystyle 1-V\frac{1}{1+K+V}} & {\displaystyle 1-V\frac{1}{1+K+V}}
\end{matrix}\right),\nonumber\\
\lineup = -\gamma B\left(1- \frac{1}{1+K+V}\right)\zeta 
+cB\left(1 -V\frac{1}{1+K+V}\right) -\gamma B\frac{1}{1+K+V} \nonumber\\
\lineup\ \ \ \ -cB\left(1- V\frac{1}{1+K+V}\right)\zeta,\nonumber\\
\lineup = 1-\zeta +\gamma \frac{B}{1+K+V}\zeta -cV\frac{B}{1+K+V}
-\gamma\frac{B}{1+K+V}+cV\frac{B}{1+K+V}\zeta,\nonumber\\
\lineup = \left(1-\gamma\frac{B}{1+K+V}-cV\frac{B}{1+K+V}\right)(1-\zeta),
\nonumber\\
\lineup=\left(1-Q\zeta \frac{B}{1+K+V}\right)(1-\zeta).
\end{eqnarray}
This matrix structure is one way to understand why this solution is so 
simple. Once we know that $g$ can be expressed as a product 
of $2\times 2$ matrices, $g^{-1}$ cannot be much more complicated.

Finally, let's verify the second part of the equations of motion,
\eq{naiveEOM}. Let's express \eq{naiveEOM} in the form 
\begin{equation}Q_{0\Psi}g=0,\end{equation}
where, following the notation of \cite{Integra}, $Q_{0\Psi}$ is the kinetic
operator for a stretched string connecting the perturbative vacuum and 
the background corresponding to the cubic solution $\Psi$:
\begin{equation}Q_{0\Psi}X = QX + 0*X - (-1)^X X\Psi.\end{equation}
Since this operator is nilpotent, the equations of motion follow from 
expressing the solution \eq{BerkVac} in the form:
\begin{equation}g = Q_{0\Psi}\left(\alpha + \zeta\frac{B}{1+K}\right).
\label{eq:QBerkVac}
\end{equation}
Here we've defined the field $\alpha\equiv-\gamma^{-2}c$, which satisfies
\begin{equation}Q\alpha = 1.\end{equation}
This corresponds to an insertion of the operator given in \eq{Qhom}. To 
see that \eq{QBerkVac} reproduces the familiar form of the solution, compute
\begin{eqnarray}g\lineup 
=Q\left(\alpha + \zeta\frac{B}{1+K}\right)+
\left(\alpha + \zeta\frac{B}{1+K}\right)\left(c-Qc\frac{B}{1+K}\right),
\nonumber\\
\lineup = 1+Q\left(\zeta\frac{B}{1+K}\right)-c\frac{B}{1+K}
+\zeta\frac{B}{1+K}c - \zeta\frac{B}{1+K}Qc\frac{B}{1+K},\nonumber\\
\lineup = 1+Q\left(\zeta\frac{B}{1+K}\right)-c\frac{B}{1+K}
+\zeta\frac{B}{1+K}c + \zeta\frac{B}{1+K}\d c\frac{1}{1+K}.
\end{eqnarray}
Now look at the $\d c$ in the last term. Write it in the form
\begin{equation}\d c = (1+K)c-c(1+K),\end{equation}
and use this to cancel the factors on either side. This gives
\begin{eqnarray}
g\lineup = 1+Q\left(\zeta\frac{B}{1+K}\right)-c\frac{B}{1+K}
+\zeta\frac{B}{1+K}c + \zeta \left(Bc\frac{1}{1+K}-\frac{B}{1+K}c\right),
\nonumber\\
\lineup = 1+ \zeta\frac{1}{1+K}+Q\left(\zeta\frac{B}{1+K}\right)
-c\frac{B}{1+K},
\end{eqnarray}
which is the solution as expressed in \eq{BerkVac2}. 

Note that any Berkovits solution can be derived from the general 
formula
\begin{equation}g=Q_{0\Psi}\beta,\label{eq:lift}\end{equation}
for the appropriate choice of cubic solution $\Psi$ and ghost number $-1$ 
field $\beta$.\footnote{The proof is as 
follows. For any Berkovits solution $g$ we can construct a cubic solution 
$\Psi=g^{-1}Qg$. Then by definition we have $Q_{0\Psi}g=0$. Further, since 
$Q\alpha =1$, we can write $g=Q_{0\Psi}(\alpha g)$.}
Since the cubic solution $\Psi$ for the superstring is often
very similar to that of the bosonic string, \eq{lift} gives an almost 
automatic lift of a solution in Witten's bosonic string field theory to a 
solution in the Berkovits superstring field theory. The challenge is choosing 
$\Psi$ and $\beta$ so that $g^{-1}$ is not too complicated or singular. 
A popular choice of $\beta$, used in many solutions for marginal 
deformations 
\cite{supermarg,Okawa_super_marg,FK_super,KO_super,simple_super_marg}, 
is $\beta=\alpha$. The tachyon vacuum \eq{BerkVacq}
comes from a slightly more complicated choice of $\beta$ which is necessary
to generate expectation values in the GSO($-$) sector. In the 
$K,B,c,\gamma,\gamma^{-1}$ subalgebra there are no nonsingular solutions 
for the tachyon vacuum using GSO($+$) states only. This is physically 
expected, and is a major advantage of the Berkovits formulation since it 
provides a clearer understanding of the role of the tachyon and the emergence 
of D-brane charges upon tachyon condensation. We discuss this further in 
section \ref{sec:charge}. 

With a short extra step we can prove the absence of open string states 
around the tachyon vacuum. Expanding around the solution \eq{BerkVac} 
gives the linearized equations of motion for a fluctuation field $\varphi$:
\begin{equation}\eta Q_\Psi\varphi = 0,\end{equation}
where $Q_\Psi$ is the shifted kinetic operator around the cubic solution 
\eq{simple}. Solutions of this equation should be identified modulo the 
linearized gauge invariance
\begin{equation}\varphi' = \varphi + Q_\Psi \Lambda+\eta\Sigma.
\end{equation}
Note that $Q_\Psi$ has a ``homotopy operator'' \cite{coh,simple}
\begin{equation}A = \frac{B}{1+K}\label{eq:hom}\end{equation}
satisfying 
\begin{equation}Q_\Psi A =1.\end{equation} 
With this we can write
\begin{equation}\varphi = Q_{\Psi}(A\varphi) + AQ_\Psi\varphi.\label{eq:nocoh}
\end{equation}
The first term is manifestly $Q_\Psi$ exact. The second term is actually $\eta$
exact, since $A$ is in the small Hilbert space and (by assumption) $\varphi$
satisfies the linearized equations of motion. Therefore 
all linearized fluctuations around the tachyon vacuum are pure gauge.

\section{Energy}
\label{sec:energy}

Now we compute the energy. To do this we compactify all directions 
(including time) tangential to the brane on circles of unit 
circumference.\footnote{This compactification is implicit in our normalization
of the correlator \eq{basic}.} Then, we can compute the energy by computing
 the action.  In our conventions, Sen's conjecture predicts 
\begin{equation}E = -S = -\frac{1}{2\pi^2}.\label{eq:tension}\end{equation}
This is minus the energy of the original unstable D-brane.

To compute the action we must choose an interpolation $g(t),t\in[0,1]$ 
connecting the identity string field to the tachyon vacuum
\eq{BerkVacq}. We choose a linear interpolation:
\begin{equation}g(t) = \bar{t}+tg,
\label{eq:lin}\end{equation}
where $\bar{t}\equiv 1-t$. This choice is convenient 
since it preserves the $2\times 2$ matrix structure of the solution, making
it possible to derive explicit expressions for both $g(t)$ and $g(t)^{-1}$:
\begin{eqnarray}
g(t)\lineup = \left(\begin{matrix}1 & qt \\ -qt & \bar{t}+K+qtV
\end{matrix}\right)\left(\begin{matrix}
1 & 0 \\ 0& {\displaystyle \frac{1}{1+K}}\end{matrix}\right),\nonumber\\
\lineup =(1+qt\zeta)\left(1+(t^2q^2-t)\,
c\frac{B}{1+K}+qt\, Q\zeta\frac{B}{1+K}
\right);\nonumber\\
g(t)^{-1}\lineup = \left(\begin{matrix}1 & 0 \\ 0 & 1+K\end{matrix}\right)
\left(\begin{matrix}\frac{\bar{t}+K+qtV}{\bar{t}+q^2t^2+K+qtV} & 
-\frac{qt}{\bar{t}+q^2t^2+K+qtV}\\ \frac{qt}{\bar{t}+q^2t^2+K+qtV} &
\frac{1}{\bar{t}+q^2t^2+K+qtV}\end{matrix}\right),\nonumber\\
\lineup = \left(1-(t^2q^2-t)\,c\frac{B}{\bar{t}+q^2t^2+K+qt\,V}
+qt\, Q\zeta \frac{B}{\bar{t}+q^2t^2+K+qt\,V}\right)(1-qt\zeta).\nonumber\\
\label{eq:linex}
\end{eqnarray}
Previous studies in Berkovits' string field theory have used the exponential 
interpolation $g(t)=e^{t\Phi}$, but this choice would substantially complicate
the analytic calculation.\footnote{Perhaps an even easier way to 
compute the energy is to compute the cubic action as in \cite{supervac}, 
and then rely on the argument of \cite{democraticgf}, based on the formalism 
of \cite{democratic}, demonstrating the on-shell equivalence of the Berkovits 
and cubic actions. We do not know of an obvious problem following the formal 
steps of \cite{democraticgf} with the solution \eq{BerkVacq}.}
Since it is not much more difficult, we compute the energy for arbitrary 
values of the gauge parameter $q$ in \eq{BerkVacq}.  

Next we compute the integrand of the action in \eq{BOZ}:
\begin{equation}\Tr[(\eta\Psi_t)\Psi_Q].\end{equation}
Plugging in $g(t)$ and $g(t)^{-1}$ produces a lengthy expression which can 
be simplified using the identities of section 
\ref{sec:algebra}. The result is
\begin{eqnarray}
\Tr[(\eta \Psi_t) \Psi_Q] \lineup =q\bar{t}\Tr\left[-\eta Q\zeta
\frac{1}{\bar{t}+q^2t^2+K+qt\,V}\right.\nonumber\\
\lineup\ \ \ \ \ \ \ \ \ \ \ \  
\left.+\eta Q\zeta\frac{2q^2t^2-t}{\bar{t}+q^2t^2+K+qt\,V}Bc
\frac{1}{\bar{t}+q^2t^2+K+qt\,V}\right.\nonumber\\
\lineup\ \ \ \ \ \ \ \ \ \ \ \  \left.+\eta Q\zeta
\frac{qt}{\bar{t}+q^2t^2+K+qt\,V}BQ\zeta\frac{1}{\bar{t}+q^2t^2+K+qt\,V}
\right].\label{eq:npsitpsiq}
\end{eqnarray}
Now expand the denominators in powers of $V$ and select the terms with 
total $bc$ ghost number $3$. With an additional reparameterization we find
\begin{eqnarray}\Tr[(\eta\Psi_t)\Psi_Q] = 
\frac{q^2 t\bar{t}}{(\bar{t}+q^2t^2)^2}X_1 
-\frac{q^2t^2\bar{t}(2q^2t-1)}{(\bar{t}+q^2t^2)^3}X_2
+\frac{q^4t^3\bar{t}}{(\bar{t}+q^2t^2)^3}X_3
+\frac{q^2t\bar{t}}{(\bar{t}+q^2t^2)^2}X_4,\nonumber\\ \label{eq:X14}
\end{eqnarray}
where
\begin{eqnarray}
X_1\lineup \equiv\Tr\left[\eta Q\zeta\frac{1}{1+K}V\frac{1}{1+K}\right];
\nonumber\\
X_2\lineup \equiv\Tr\left[\eta Q\zeta\frac{B}{1+K}V\frac{1}{1+K}c\frac{1}{1+K}
\right]+\Tr\left[\eta Q\zeta\frac{B}{1+K}c\frac{1}{1+K}V\frac{1}{1+K}\right];
\nonumber\\
X_3\lineup \equiv\Tr\left[\eta Q\zeta \frac{B}{1+K}V\frac{1}{1+K}V\frac{1}{1+K}
\gamma\frac{1}{1+K}\right]
+\Tr\left[\eta Q\zeta \frac{B}{1+K}V\frac{1}{1+K}\gamma\frac{1}{1+K}
V\frac{1}{1+K}\right]\nonumber\\
\lineup\ \ \ \ \ \ \ \ \ \ \ \ \ \ 
+\Tr\left[\eta Q\zeta \frac{B}{1+K}\gamma \frac{1}{1+K}V\frac{1}{1+K}
V\frac{1}{1+K}\right];\nonumber\\
X_4\lineup \equiv\Tr\left[\eta Q\zeta\frac{B}{1+K}cV\frac{1}{1+K}\right].
\end{eqnarray}
$X_1,...,X_4$ are simply constants which we can compute by evaluating the 
respective worldsheet correlation functions. We will do this in appendix 
\ref{app:energy}. The result is
\begin{eqnarray}
X_1\lineup = 0;\nonumber\\
X_2\lineup = -\frac{1}{\pi^2};\nonumber\\
X_3\lineup = 0;\nonumber\\
X_4\lineup = -\frac{2}{\pi^2}.\label{eq:Xs}
\end{eqnarray}
Plugging into \eq{X14} gives
\begin{equation}\Tr[(\eta\Psi_t)\Psi_Q] = 
-\frac{1}{\pi^2}\left( -\frac{q^2t^2\bar{t}(2q^2t-1)}{(\bar{t}+q^2t^2)^3}
+\frac{2q^2t\bar{t}}{(\bar{t}+q^2t^2)^2}\right).\label{eq:integrand}
\end{equation}
To find the energy we integrate from $0$ to $1$:
\begin{equation}E = -S = \int_0^1 dt\Tr[(\eta\Psi_t)\Psi_Q].\end{equation}
To compute this integral, write the $q^2t^2$ factor in the 
numerator of the first term of \eq{integrand} in the form
\begin{equation}q^2t^2 = (\bar{t}+q^2t^2)-\bar{t},\end{equation}
and cancel with the denominator:
\begin{eqnarray}\Tr[(\eta\Psi_t)\Psi_Q] \lineup = 
-\frac{1}{\pi^2}\left( -
\frac{((\bar{t}+q^2t^2)-\bar{t})\bar{t}(2q^2t-1)}{(\bar{t}+q^2t^2)^3}
+\frac{2q^2t\bar{t}}{(\bar{t}+q^2t^2)^2}\right),\nonumber\\
\lineup = -\frac{1}{\pi^2}\left( 
\frac{\bar{t}^2(2q^2t-1)}{(\bar{t}+q^2t^2)^3}
+\frac{\bar{t}}{(\bar{t}+q^2t^2)^2}\right).
\end{eqnarray}
Now note that the quantity $2q^2 t-1$ in the numerator of the first term
is the derivative of $\bar{t}+q^2t^2$ in the denominator. Thus it is easy
to see that 
\begin{equation}
\Tr[(\eta\Psi_t)\Psi_Q] = -\frac{1}{\pi^2}\frac{d}{dt}\left(-\frac{1}{2}
\frac{\bar{t}^2}{(\bar{t}+q^2t^2)^2}\right),
\end{equation}
and
\begin{eqnarray}E = \frac{1}{2\pi^2}\left.
\frac{\bar{t}^2}{(\bar{t}+q^2t^2)^2}\right|_{t=0}^{t=1} = -\frac{1}{2\pi^2}.
\end{eqnarray}
in agreement with Sen's conjecture.

Another way to detect the energy is to probe the solution with a closed 
string. This can be accomplished by computing the Ellwood invariant 
\cite{Ellwood}, which is believed to describe the shift in the closed string 
tadpole amplitude between the perturbative vacuum and the background 
described by a classical solution.\footnote{A formal argument relating the 
Ellwood invariant and the value of the on-shell action was given in 
\cite{Ishibashi} for the bosonic string. It would be interesting to extend 
this argument to the superstring.} In Berkovits' string field theory, 
the Ellwood invariant comes in three varieties:
\begin{eqnarray}
Q\mbox{-Ellwood\ Invariant} \lineup = \Tr_{V_Q}\!
\left[\!\left.\Psi_Q\right|_{t=1}\right];\label{eq:QEll}\\
t\mbox{-Ellwood\ Invariant} \lineup = \int_0^1 dt\, \Tr_{V_t}[\Psi_t],
\label{eq:tEll}\\
\eta\mbox{-Ellwood\ Invariant} \lineup = \Tr_{V_\eta}\!
\left[\!\left.\Psi_\eta\right|_{t=1}\right];\label{eq:etaEll}
\end{eqnarray} 
where $\Tr_V[\cdot]$ is the trace with a midpoint insertion of the vertex 
operator $V$, and the vertex operators in each case are
\begin{equation}V_Q = \xi V_{-2},\ \ \ V_t = V_{-1},\ \ \ V_\eta = \alpha V_0.
\end{equation}
Here $V_{-2},V_{-1}$ and $V_0$ are on-shell closed string vertex operators 
in the $-2,-1$ and $0$ picture respectively, killed by $\eta$ and $Q$
and with vanishing conformal dimension. The operators $\xi$ and $\alpha$ 
above are some combination of the left/right zero modes of $\xi$ and $\alpha$ 
in \eq{Qhom}.\footnote{The triplet of Ellwood invariants is reminiscent of the 
triplet of cubic, WZW-like, and dual cubic actions observed in 
\cite{democraticgf}.} Though the three Ellwood invariants look different, 
they compute the same quantity. For example, assuming $V_t = QV_Q$ we have 
\begin{eqnarray}
\int_0^1 dt\, \Tr_{V_t}[\Psi_t]\lineup = \int_0^1 dt\, \Tr_{V_Q}[Q\Psi_t],
\nonumber\\
\lineup = \int_0^1 dt\, \Tr_{V_Q}[\d_t'\Psi_Q],\nonumber\\
\lineup = \int_0^1 dt\, \d_t \Tr_{V_Q}[\Psi_Q],\nonumber\\
\lineup = \Tr_{V_Q}\!\left[\!\left.\Psi_Q\right|_{t=1}\right].
\end{eqnarray}
Therefore it is enough to compute the $Q$-Ellwood invariant, for which we 
don't really need the Berkovits solution---the cubic solution 
\eq{simple} is sufficient. Then the computation reduces to that of the 
bosonic string \cite{simple}, reproducing the expected closed string tadpole 
amplitude of the reference D-brane. The $t$- and 
$\eta$-invariants will compute the same amplitude, but their first 
quantized interpretation will be different since the closed string vertex 
operator lives in a different picture. It would be interesting to compute 
these invariants and clarify their first quantized interpretation. 

\section{Tachyon Coefficient}
\label{sec:coefficient}

It is useful to consider the Fock space expansion of the solution 
\eq{BerkVacq}, both for the purpose of comparison with earlier numerical 
solutions \cite{BSZ,Raey,Iqbal} and in general to understand the solution's 
properties in level truncation. Though we are not able to execute a high level 
analysis, as a first step we can compute the tachyon coefficient $T$
\begin{equation}T\, i\sigma_1\otimes \xi c e^{-\phi}(0)|0\rangle,
\label{eq:FockTach}\end{equation}
which represents the expectation value of the tachyon field at the tachyon 
vacuum. 

Specifically, we want to compute the tachyon coefficient of the Lie algebra 
element
\begin{equation}\Phi=\ln g.\end{equation}
Since the solution is a product of two noncommuting $2\times 2$ matrices, 
in principle we can compute $\Phi$ by taking the logarithm of each matrix 
factor and substituting into the Campbell-Baker-Hausdorff formula. But there 
is a simpler way to do this. Consider an interpolation 
$g(t)$ which can be written as a function of the solution $g$ and the 
parameter $t$ only:
\begin{equation}g(t) = f(t,g).\end{equation}
We call this an {\it abelian} interpolation. Assuming $g(t)$ is abelian,
we can compute $\Phi$ with the formula\footnote{This implies that the 
$t$-Ellwood invariant can be computed as $\Tr_{V_t}[\Phi]$, as described 
by Michishita \cite{Michishita}. However, the expression \eq{tEll} is more 
general since it does not assume an abelian interpolation.}
\begin{equation}\Phi = \int_0^1 dt\, \Psi_t.\label{eq:intPhi}\end{equation}
The proof is as follows. Let $\delta$ represent a variation relating abelian 
interpolations with fixed boundary conditions at $t=0$ and $t=1$. Then
\begin{equation}\delta \int_0^1 dt\, \Psi_t = \int_0^1 dt\,
\big(\d_t\Psi_\delta+[\Psi_t,\Psi_\delta]\big),\end{equation}
where we used \eq{vFT}. Since the variation preserves the abelian property, 
$\Psi_t$ and $\Psi_\delta$ are functions of $g$ and $t$ only,
and commute. The integral of the total derivative vanishes 
since by assumption the variation vanishes at $t=0$ and $1$. Therefore
\begin{equation}\delta\int_0^1 dt\,\Psi_t = 0,\end{equation}
which means that the integral \eq{intPhi} is independent of the
choice of abelian $g(t)$. Then choosing $g(t)=e^{t\Phi}$ establishes 
the result.

In particular, the linear interpolation \eq{lin} is an abelian interpolation. 
Therefore substituting \eq{linex} into \eq{intPhi} gives a 
formula for the Lie algebra element of the tachyon vacuum solution 
\eq{BerkVacq}:
\begin{eqnarray}\Phi \lineup = q\zeta 
+ cB\left(\int_0^1 dt\, \frac{q^2t-1}{\bar{t}+q^2t^2+K+qt V}\right)
-cB\left(\int_0^1 dt\, \frac{qt(q^2t-1)}{\bar{t}+q^2t^2+K+qt V}\right)\zeta
\nonumber\\
\lineup\ \ \ \ \ \ \ +Q\zeta B \left(\int_0^1 dt\, 
\frac{q}{\bar{t}+q^2t^2+K+qt V}\right)-Q\zeta B \left(\int_0^1 dt\, 
\frac{q^2t}{\bar{t}+q^2t^2+K+qt V}\right)\zeta.\nonumber\\
\label{eq:BerkVac_Phi}
\end{eqnarray}
We cannot easily integrate over $t$ in this formula since the integrand is 
noncommutative. However, we do not really need to perform 
these integrals. We simply contract with a Fock space state and leave the 
integration over $t$ as a final step once the integrand has turned into an 
ordinary function.

Now we compute the tachyon coefficient. To do this, we contract $\Phi$ with 
the test state dual to the zero momentum tachyon\footnote{The normalization
is chosen so that the BPZ inner product of test state \eq{test} and the 
Fock space zero momentum tachyon \eq{FockTach} is the tachyon coefficient 
$T$. Since the dual test state has $L_0=-\frac{1}{2}$ it is orthogonal to 
all other Fock space states. A more efficient method for extracting 
Fock space coefficients at higher levels would utilize the operator formalism
of Schnabl \cite{Schnabl}.}
\begin{equation}
 -\sqrt{\frac{\pi}{2}}\,\sqrt{\Omega}(\eta\gamma^{-1}c\d c)
\sqrt{\Omega}.\label{eq:test}
\end{equation}
So the tachyon coefficient is given by
\begin{equation}T = -\sqrt{\frac{\pi}{2}}
\Tr\Big[\sqrt{\Omega}(\eta\gamma^{-1}c \d c)\sqrt{\Omega}\,\Phi\Big].
\end{equation}
With the help of the correlators in appendix \ref{app:correlators} we
find the result
\begin{eqnarray}
T\lineup =\frac{q}{\sqrt{2\pi}}\left[1+\int_0^\infty dL\int_0^1 dt\,
e^{-L(1-t+q^2t^2)}\left((t+1 -(1+L)q^2t^2)\sin\frac{\pi}{2(1+L)}
\right.\right.\nonumber\\
\lineup\ \ \ \ \ \ \ \ \ \ \ \ \ \ \ \ \ \ \ \ \ \ \ \ \ \ \ \ \ \ \ \ \ \ \ 
\ \ \ \ \ \ \ \ \ \ \ \ \ \ \ \ \ \ 
\left.\left.-\frac{\pi}{2(1+L)}\cos\frac{\pi}{2(1+L)}\right)\right].\label{eq:tachcoef}
\end{eqnarray}
Here $1+L$ is the circumference of the cylinder obtained upon expanding the 
Lie algebra element \eq{BerkVac_Phi} in terms of wedge states. The 
integration over $t$ can be performed analytically in terms of error 
functions, but makes the formula look more complicated.

\begin{figure}
\begin{center}
\resizebox{4.75in}{3in}{\includegraphics{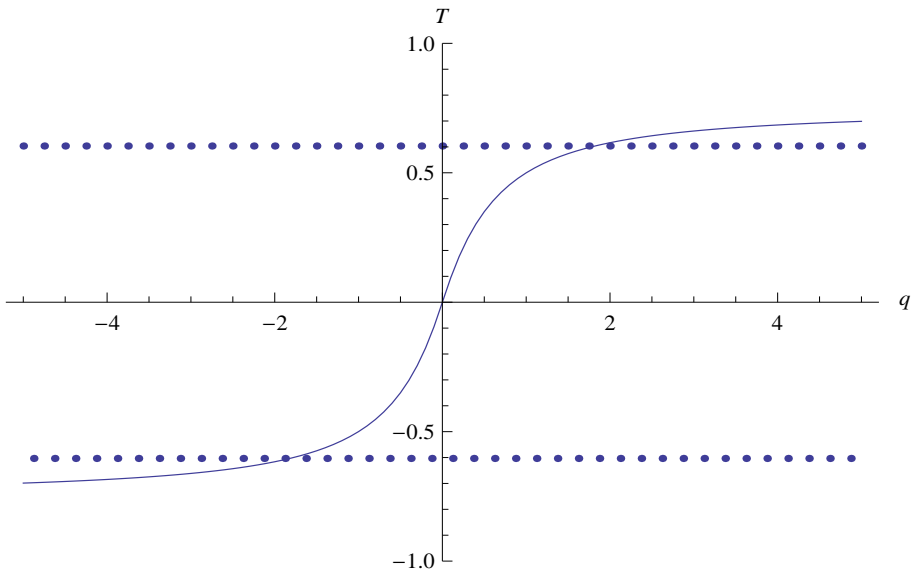}}
\end{center}
\caption{\label{fig:BerkVac9} Tachyon coefficient plotted as a function of 
$q\in[-5,5]$. The dotted lines represent the current best approximation to 
the tachyon coefficient of the Siegel-$(\xi_0=0)$ gauge condensate,
$T\approx \pm .615$ \cite{DeSmet}.}
\end{figure}

Thus we have determined the tachyon coefficient as a function of the gauge 
parameter $q$. We have plotted this in figure \ref{fig:BerkVac9}. The tachyon 
coefficient is an odd, monotonically increasing function of $q$. For $q$ not 
too small, the coefficient is in the ball-park of its approximate value 
in the Siegel-$(\xi_0=0)$ gauge condensate, $T\approx .6$. At $q=1$ the 
tachyon coefficient is
\begin{equation}T|_{q=1}\approx .4998 \end{equation}
which is surprisingly close to value $T=\frac{1}{2}$ computed
from the action truncated to level zero \cite{BerkTrunc}. When 
$q\to\pm\infty$ the tachyon approaches a finite upper/lower bound
\begin{equation}\lim_{q\to\pm\infty}T\approx \pm \,.7554.
\end{equation}
For small $q$ the tachyon coefficient approaches zero. Though $q\to 0$ looks 
smooth, this is a very singular limit of the solution. This is 
expected since the process of tachyon condensation should generate an 
expectation value  for the tachyon.\footnote{The absence of a tachyon in the 
cubic solution \eq{simple} has been somewhat of a puzzle, and probably 
indicates that the cubic equations of motion do not accurately capture the 
nonperturbative solution space of the superstring \cite{exotic}.}

\section{D-brane Charge?}
\label{sec:charge}

The tachyon effective potential should have a pair of global minima
corresponding to two tachyon vacuum solutions,
\begin{equation}g,\ g',\end{equation}
related by a sign flip in the GSO($-$) sector:
\begin{equation} g'=(-1)^F g.\end{equation}
These two solutions are gauge equivalent. For the analytic solution 
\eq{BerkVacq}, the finite gauge transformation relating them is simply
\begin{equation}g' = (g'g^{-1})g,\end{equation}
since the product $g'g^{-1}$ is BRST closed.\footnote{In the general situation,
the solutions $g$ and $g'$ correspond to different cubic solutions, and 
the gauge transformation relating them requires an $\eta$ closed 
factor as well.} Therefore $g$ and $g'$ should represent the same physical 
state, even though they have opposite GSO($-$) expectation values. 

But this raises a puzzle. A non-BPS D-brane of Type II should have a 
codimension 1 kink solution interpolating between $g$ and $g'$ describing 
a stable BPS D-brane of one lower dimension.\footnote{Some attempts to 
derive analytic solutions for lower dimensional branes are described in 
\cite{Ian_proj,BMT,lumps,Calcagni}.} But since  $g$ and $g'$ are 
physically equivalent, one might think that the kink solution is really 
a lump in disguise, and could dissipate. Turning the kink into a lump 
requires a gauge transformation which acts as the identity on one end 
of the kink, and turns $g$ into $g'$ on the other end. If $x\in\mathbb{R}$ 
is the coordinate along the kink, we need a gauge parameter $V(x)$ satisfying
\begin{equation}\lim_{x\to-\infty}V(x) = 1,\ \ \ \ \lim_{x\to\infty}V(x) = 
g'g^{-1}.\label{eq:homV}\end{equation}
Thus $V(x)$ would define a homotopy between $g'g^{-1}$ and the identity 
string field. We suggest that no such homotopy exists, which is why the 
kink solution is stable. Thus $g'g^{-1}$ is a topologically nontrivial, 
``large'' gauge transformation.

This is a fairly ambitious statement. 
We will not try to prove it, but instead provide some evidence based on 
our analytic solution \eq{BerkVacq}. 
The existence of $V(x)$ in \eq{homV} would imply the existence of a homotopy 
$g(\lambda),\lambda\in[0,1]$ of tachyon vacuum solutions connecting 
$g$ and $g'$:
\begin{equation}g(0)=g;\ \ \ g(1)=g'.\end{equation}
We can look for this homotopy within the class of analytic solutions we have 
studied, that is, assuming solutions of the form \eq{BerkVacq} 
which differ only in the choice of $q$. Thus $g(\lambda)$
corresponds to $q(\lambda),\lambda\in[0,1]$ satisfying
\begin{equation}q(0)=q;\ \ \ q(1)=-q.\end{equation}
We claim that $g(\lambda)$ must be singular for at least one $\lambda$. To 
see this, note that $g(\lambda)^{-1}$ involves the state
\begin{equation}\frac{1}{q(\lambda)^2+K}.\label{eq:probstate}\end{equation}
According to a proposal of Rastelli \cite{Rastelli}, the algebra of wedge 
states should correspond to the $C^*$-algebra of bounded, continuous 
functions on the positive real line $K\geq 0$. Applying this criterion to 
\eq{probstate} implies that $q(\lambda)$ cannot be zero or imaginary. 
But there is no path connecting $q$ to $-q$ which does not pass through 
the imaginary axis. Therefore, at least within the class of gauge 
transformations preserving \eq{BerkVacq}, $g'g^{-1}$ is not homotopic to 
the identity. This is consistent with stability of the kink.

\begin{figure}
\begin{center}
\resizebox{4in}{2.6in}{\includegraphics{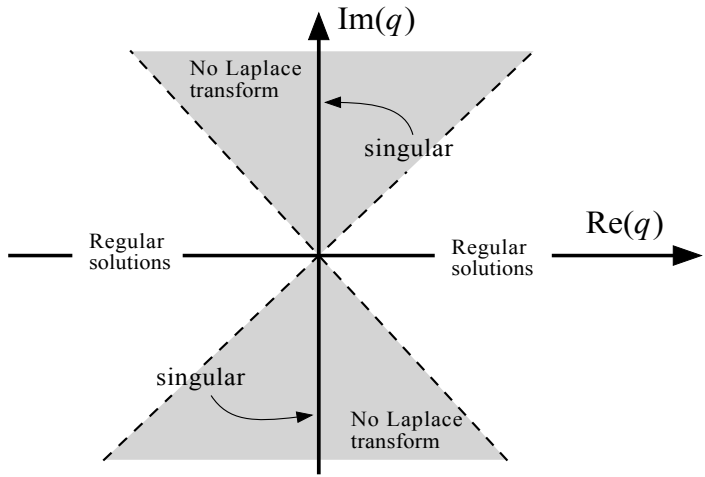}}
\end{center}
\caption{\label{fig:BerkVac11} Space of tachyon vacuum solutions of the form
\eq{BerkVacq} in the complex $q$ plane. For purely imaginary $q$ the solutions
are singular, and in the shaded region they are not definable in terms of 
superpositions of wedge states. Thus the solution space has two disconnected
components corresponding to the two minima of the tachyon effective
potential.}
\end{figure}

Actually, the state \eq{probstate} arguably becomes singular long before 
$q(\lambda)$ reaches the imaginary axis. Currently, the only practical way 
for defining states in the wedge algebra is as a Laplace transform 
\begin{equation}F(K)=\int_0^\infty d\alpha\,f(\alpha)\Omega^\alpha.
\end{equation}
This representation assumes that $F(K)$ is analytic on the positive half of 
the complex plane $\mathrm{Re}(K)>0$. Applying this to \eq{probstate} implies
\begin{equation}|\mathrm{Re}(q)|>|\mathrm{Im}(q)|.\end{equation}
This excludes not only the imaginary axis, but also a cone of solutions around
the imaginary axis as shown in figure \ref{fig:BerkVac11}. It is possible 
that some solutions inside the cone could be understood with some more 
general definition of the algebra of wedge states \cite{exotic}, but this is 
unclear. 

\begin{figure}
\begin{center}
\resizebox{3in}{2.7in}{\includegraphics{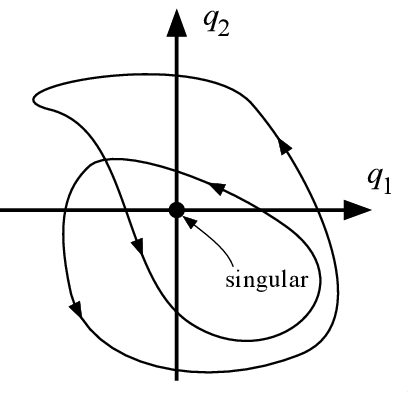}}
\end{center}
\caption{\label{fig:BerkVac12} Space of tachyon vacuum solutions of the form 
\eq{BerkVacq} on a brane-antibrane pair, for Hermitian 
$q=q_1\sigma_1+q_2\sigma_2$. The only singular solution is at 
$q_1=q_2=0$. Removing this point allows homotopically nontrivial 
windings around the gauge orbit, which should represent the charges
of BPS D-branes of codimension 2 on the brane-antibrane pair.}
\end{figure}

One might ask why this argument does not also imply stability of the kink on 
the brane-antibrane pair, which should represent an unstable non-BPS D-brane.
The solution \eq{BerkVacq} can be immediately generalized to the 
brane-antibrane by taking $q$ to be an arbitrary off-diagonal $2\times 2$ 
matrix
\begin{equation}q = q_1\sigma_1+q_2\sigma_2.\end{equation}
This off-diagonal matrix represents the (external) Chan-Paton factors for 
the GSO($-$) strings connecting the brane and antibrane. Again we look for a 
homotopy $g(\lambda),\lambda\in[0,1]$ connecting $g$ and $g'$ within the ansatz
\eq{BerkVacq}. This time, the relevant state which appears in $g(\lambda)^{-1}$
is
\begin{equation}\frac{1}{q_1(\lambda)^2+q_2(\lambda)^2+K}.\end{equation}
Restricting for simplicity to real $q_1,q_2$, the only problematic point for
this state is $q_1=q_2=0$. This point is easily avoided in a path from $q$ to 
$-q$. Thus on the brane-antibrane pair $g'g^{-1}$ is homotopic to the 
identity, and the kink solution is unstable. Incidentally, note that the 
``hole'' at $q_1=q_2=0$ suggests the existence of codimension 2 topological 
solitons obtained by winding around the tachyon vacuum gauge orbit at 
infinity (see figure \ref{fig:BerkVac12}). These solitons are the expected 
BPS D-($p-2$)-branes formed by tachyon condensation on the 
D$p$-$\bar{\mathrm{D}}p$ system. We expect that higher codimension brane 
charges can be seen by looking at the tachyon vacuum on multiple non-BPS 
D-branes or brane-antibrane pairs. The picture so far seems consistent 
with the intuition for D-brane charge derived from the presumed form of the 
tachyon effective potential.

Understanding the topological charge of stable D-branes is a long standing 
and fundamental problem in string field theory.  But we have only looked at 
a small slice of the gauge orbit, and we need to demonstrate that the 
topological structures we've observed survive a more general analysis. 
With further development, we hope that it is possible to strengthen our 
argument based on analysis of the full tachyon vacuum gauge orbit in the
 $K,B,c,\gamma,\gamma^{-1}$ subalgebra.\footnote{A 
toy model for this kind of problem appears in the analysis of the gauge 
orbit of ``half brane'' solutions in cubic superstring field theory 
\cite{exotic}.} Another question is how this characterization of D-brane 
charge leads to Ramond-Ramond flux. Analysis of the boundary state 
\cite{KOZ,KMS} or the Ellwood invariant \cite{Ellwood} may provide insight 
into this question. It would also be interesting to see if these 
developments can shed light on the long-speculated relation between string 
field theory and the $K$-theoretic description of D-brane charge 
\cite{Moore,WittenK,WittenK2}. We leave these questions for future work.

\section{Conclusion}

In this paper we found an exact solution for tachyon 
condensation in Berkovits' nonpolynomial open superstring field theory.
The solution is completely explicit and very simple. The main obstacle to 
finding the solution was not necessarily the equations of motion---a simple 
strategy for solving the equations of motion has been known since
\cite{supermarg,FK_super,FKcub}. The main obstacle was finding a 
specific solution such that the various nonpolynomial expressions needed 
in the theory do not produce an uncontrolled proliferation of superghost 
insertions. Our main innovations in this respect were the introduction 
of the boundary interaction \eq{Vins} in conjunction with the $2\times 2$ 
matrix algebra \eq{2by2}. With these ingredients we have been able to 
whittle the computation of the action down to a single correlator with 
merely two superghost insertions. Thus we are able to explicitly prove 
Sen's conjecture and provide the first nontrivial analytic computation 
of the nonpolynomial action. 

The solution opens a number of interesting avenues for exploration. The most
immediate is obtaining new solutions and a clearer understanding of the 
tachyon vacuum gauge orbit and its relation to D-brane charge. Next, it 
is desirable to incorporate the Ramond sector and understand the role of 
supersymmetry. In principle, it should be possible to derive BPS equations 
from the spontaneously broken supersymmetries on a non-BPS D-brane, and 
solve these equations to find lower dimensional BPS D-branes as topological 
solitons. Another urgent question is the role of closed strings. While the 
perturbative quantization of the Berkovits theory is not yet well understood, 
this is a topic of active research \cite{BerkPert,BerkPertBerk,Torii1,Torii2}.
We hope that the tachyon vacuum solution stimulates progress in these and 
other important problems.

\bigskip

\noindent {\bf Acknowledgments}

\bigskip

\noindent The author would like to thank M. Schnabl for collaboration in an 
earlier stage of this project and for detailed comments and corrections on 
the second version of this manuscript. The author also thanks Y. Okawa 
for hospitality at the University of Tokyo where the solution was 
discovered, S. Hellerman for valuable conversations, and C. Maccaferri 
for comments on the manuscript. This research was supported by the Grant 
Agency of the Czech Republic under the grant P201/12/G028.

\begin{appendix}

\section{Correlators}
\label{app:correlators}

Here we list some correlators which are needed for our calculations. The 
correlators are normalized according to the convention,
\begin{equation}\langle\xi(z)c\d c\d^2c(w)e^{-2\phi}(y)\rangle
\equiv 2.\end{equation}
We will give formulas for correlation functions on the cylinder of 
circumference $L$, which following \cite{Okawa} we denote with a subscript 
$C_L$. These are related to correlation functions on the upper half plane
through the conformal transformation
\begin{equation}\langle ... \rangle_{C_L} = 
\left\langle f_\mathcal{S}^{-1}\circ\frac{2}{L}\circ(...)
\right\rangle_{UHP},\end{equation}
where $f_\mathcal{S}^{-1}(z)=\tan \frac{\pi z}{2}$ is the inverse of the 
sliver coordinate map \cite{RZ} and $\frac{2}{L}$ is a dilatation by a factor
of $\frac{2}{L}$:
\begin{equation}
f_\mathcal{S}^{-1}\circ\frac{2}{L}(z)=\tan \frac{\pi z}{L}.
\label{eq:confL}\end{equation}
The most general correlator in the $K,B,c,\gamma,\gamma^{-1}$ subalgebra is
\begin{equation}\left\langle\phantom{\frac{A}{B}}\!\!\!\!\!\!
\Big(\eta\gamma^{-1}(w)
\gamma^{-1}(x_1)\gamma^{-1}(x_2)\, .\, .\, .\, \gamma^{-1}(x_{n+1}) \gamma(y_1)
\gamma(y_2)\, .\, .\, .\, \gamma(y_n)\Big)\Big(c(z_1)c(z_2)c(z_3)c(z_4)B\Big)
\right\rangle_{C_L}.
\label{eq:gen_corr}\end{equation}
Here we have already striped away and traced over internal CP factors. 
Without loss of generality, we can assume that the coordinates $w,x_i,y_i$ and 
$z_i$ sit on the real axis between $\mathrm{Re}(z)=L$ and $\mathrm{Re}(z)=0$, 
since the conformal transformation \eq{confL} identifies coordinates outside 
this range modulo $z\sim z+L$. This is what it means for \eq{gen_corr} to 
be a correlation function on the cylinder. We also assume that the $B$ 
contour meets the real axis somewhere between $0$ and the smallest
$z_i$ representing an insertion of $c$.

The correlator \eq{gen_corr} can be computed as a product of correlators 
in the $\beta\gamma$ and $bc$ conformal field theories. The $\beta\gamma$ 
factor is
\begin{eqnarray}\lineup \!\!\!\!\!\!\!\!\!\!
\Big\langle \eta\gamma^{-1}(w)  
\gamma^{-1}(x_1)\gamma^{-1}(x_2)\, .\, .\, .\, \gamma^{-1}(x_{n+1})\gamma(y_1)
\gamma(y_2)\, .\, .\, .\, \gamma(y_n)\Big\rangle_{C_L}^{\beta\gamma}\nonumber\\
\lineup \ \ \ \ \ \ \ \ \ \ \ \ \ \ \ \ \ \ \ \ \ \ \ \ \ \ \ \ \ 
\ \ \ \ \ \ \ \ \ \ \ \ \ \ \ \ \ \ \ \ \ \ \ \ \ \ \  
=\frac{\pi}{L}\,\frac{\sin\frac{\pi(w-y_1)}{L} ... 
\sin\frac{\pi(w-y_n)}{L}}{\sin\frac{\pi(w-x_1)}{L}\sin\frac{\pi(w-x_2)}{L}
...\sin\frac{\pi(w-x_{n+1})}{L}}.\nonumber\\
\label{eq:gginv}
\end{eqnarray}
The $bc$ factor is \cite{Schnabl,Okawa}
\begin{eqnarray}\langle c(z_1)c(z_2)c(z_3)c(z_4)B\rangle^{bc}_{C_L}
\lineup =\frac{L^2}{\pi^3}\left(
z_1\sin\frac{\pi z_{23}}{L}\sin\frac{\pi z_{24}}{L}\sin\frac{\pi z_{34}}{L}
-z_2\sin\frac{\pi z_{13}}{L}\sin\frac{\pi z_{14}}{L}\sin\frac{\pi z_{34}}{L}
\right.\nonumber\\
\lineup \ \ \ \ \ \ \left.
+z_3\sin\frac{\pi z_{12}}{L}\sin\frac{\pi z_{14}}{L}\sin\frac{\pi z_{24}}{L}
-z_4\sin\frac{\pi z_{12}}{L}\sin\frac{\pi z_{13}}{L}\sin\frac{\pi z_{23}}{L}
\right),\nonumber\\
\label{eq:bcnew}
\end{eqnarray}
where $z_{ij}\equiv z_i-z_j$. Another useful form is 
\begin{eqnarray}\langle c(z_1)c(z_2)c(z_3)c(z_4)B\rangle^{bc}_{C_L}
\lineup =\frac{L^2}{4\pi^3}\left(z_{14}\sin\frac{2\pi z_{23}}{L}
+z_{23}\sin\frac{2\pi z_{14}}{L}\right.\nonumber\\
\lineup \ \ \ \ \ \ \ \ \ \ \ 
\left.-z_{13}\sin\frac{2\pi z_{24}}{L}
-z_{24}\sin\frac{2\pi z_{13}}{L}\right.\nonumber\\
\lineup \ \ \ \ \ \ \ \ \ \ \ 
\left.+z_{12}\sin\frac{2\pi z_{34}}{L}
+z_{34}\sin\frac{2\pi z_{12}}{L} \right).
\label{eq:bcold}
\end{eqnarray}
We also list a few special cases which appear in our calculations:
\begin{eqnarray}
\langle c\d c(z_1) c(z_2)\rangle_{C_L}^{bc} \lineup 
= -\left(\frac{L}{\pi}\right)^2 \sin^2\frac{\pi z_{12}}{L}; \\
\langle c\d c(z_1) \d c(z_2)\rangle_{C_L}^{bc} \lineup 
=\frac{L}{\pi}\sin\frac{2\pi z_{12}}{L};\\
\langle c\d c(z_1) c(z_2) c(z_3) B\rangle_{C_L}^{bc} \lineup 
=\frac{L^2}{\pi^3}\left[\frac{\pi}{L}\left(z_{12}\sin^2\frac{\pi z_{13}}{L}-
z_{13}\sin^2\frac{\pi z_{12}}{L}\right)-\sin\frac{\pi z_{12}}{L}
\sin\frac{\pi z_{13}}{L} \sin\frac{\pi z_{23}}{L}\right];\nonumber\\
\\
\langle c\d c(z_1) \d c(z_2) c(z_3) B\rangle_{C_L}^{bc} \lineup
=\frac{2L}{\pi^2}\sin\frac{\pi z_{12}}{L}\left(\frac{\pi z_{13}}{L}
\cos\frac{\pi z_{12}}{L}-\cos\frac{\pi z_{23}}{L}\sin\frac{\pi z_{13}}{L}
\right) ;\\
\langle c\d c(z_1) \d c(z_2) \d c(z_3) B\rangle_{C_L}^{bc} \lineup
=-\frac{4}{\pi}\sin\frac{\pi z_{12}}{L}\sin\frac{\pi z_{13}}{L}
\sin\frac{\pi z_{23}}{L};\\
\langle c\d c(z_1)c\d c(z_2) B\rangle_{C_L}^{bc}\lineup =
\frac{2L}{\pi^2}\sin\frac{\pi z_{12}}{L}\left(\sin\frac{\pi z_{12}}{L}
-\frac{\pi z_{12}}{L}\cos\frac{\pi z_{12}}{L}\right).\label{eq:bcenergy}
\end{eqnarray}
We use these correlators to compute the tachyon coefficient in \eq{tachcoef}.
Only \eq{bcenergy} appears in the computation of the action.

\section{Energy Coefficients}
\label{app:energy}

In this appendix we calculate the constants $X_1,...,X_4$ which appear in 
the computation of the action. 

The constants $X_1$ and $X_3$ vanish. Actually we can prove this 
without calculating any correlators. For $X_1$ we can see this as follows: 
\begin{eqnarray}
X_1\lineup = \Tr\left[\eta Q \zeta \frac{1}{1+K}[B,Q\zeta]\frac{1}{1+K}\right],
\nonumber\\
\lineup =-\Tr\left[[B,\eta Q \zeta] \frac{1}{1+K}Q\zeta\frac{1}{1+K}\right],
\nonumber\\
\lineup =\Tr\left[\eta V \frac{1}{1+K}Q\zeta\frac{1}{1+K}\right],
\nonumber\\
\lineup =-\Tr\left[ V \frac{1}{1+K}\eta Q\zeta\frac{1}{1+K}\right],
\nonumber\\
\lineup = -\Tr\left[ \eta Q\zeta \frac{1}{1+K}V\frac{1}{1+K}\right] = -X_1.
\end{eqnarray}
Meanwhile, for $X_3$
\begin{eqnarray}
X_3\lineup =\Tr\left[\eta Q\zeta \frac{B}{1+K}Q\zeta\frac{B}{1+K}Q\zeta
\frac{B}{1+K}\gamma\frac{1}{1+K}\right]
+\Tr\left[\eta Q\zeta \frac{B}{1+K}Q\zeta\frac{B}{1+K}\gamma\frac{1}{1+K}
Q\zeta\frac{B}{1+K}\right]\nonumber\\
\lineup\ \ \ \ \ \ \ \ \ \ \ \ \ \ 
+\Tr\left[\eta Q\zeta \frac{B}{1+K}\gamma \frac{1}{1+K}Q\zeta\frac{B}{1+K}
Q\zeta\frac{B}{1+K}\right],\nonumber\\
\lineup = \Tr\left[\eta Q\zeta \frac{B}{1+K}Q\zeta\frac{B}{1+K}Q\zeta
\frac{B}{1+K}\gamma\frac{1}{1+K}\right]
+\Tr\left[ Q\zeta \frac{B}{1+K}\eta Q\zeta\frac{B}{1+K}Q\zeta\frac{B}{1+K}
\gamma\frac{1}{1+K}\right]\nonumber\\
\lineup\ \ \ \ \ \ \ \ \ \ \ \ \ \ 
+\Tr\left[Q\zeta \frac{B}{1+K}Q\zeta \frac{B}{1+K}\eta Q\zeta\frac{B}{1+K}
\gamma\frac{1}{1+K}\right],\nonumber\\
\lineup =\Tr\left[\eta\left(Q\zeta \frac{B}{1+K}Q\zeta \frac{B}{1+K}Q\zeta\frac{B}{1+K}
\gamma\frac{1}{1+K}\right)\right] = 0.
\end{eqnarray}

Now let's compute $X_4$. To match with correlators given in 
appendix \ref{app:correlators}, it is useful to move the $B$ insertion in 
$X_4$ to the right. Commuting the $B$ past the $c$ produces a term 
proportional to $X_1$, which vanishes as just demonstrated. Therefore 
we write $X_4$:
\begin{equation}
X_4 = -\Tr\left[\eta Q\zeta\frac{1}{1+K}cV\frac{B}{1+K}\right].
\label{eq:X4new}
\end{equation}
Next we compute:
\begin{eqnarray}
X_4 \lineup = -\int_0^\infty d\alpha_1d\alpha_2\, e^{-(\alpha_1+\alpha_2)}
\, \Tr\Big[(\eta Q\zeta)  \Omega^{\alpha_1}(cV)
B\Omega^{\alpha_2}\Big],\nonumber\\
\lineup = -\int_0^\infty dL\,Le^{-L}\int_0^1 d\theta 
\Tr\Big[(\eta Q\zeta)  \Omega^{L\theta}(cV)B\Omega^{L(1-\theta)}\Big],
\nonumber\\
\lineup = -\int_0^1 d\theta\, 
\Tr\Big[(\eta Q\zeta) \Omega^{\theta}(cV)B \Omega^{1-\theta}\Big].
\label{eq:expand}
\end{eqnarray}
In the first step we expanded the two factors of $\frac{1}{1+K}$ into integrals
over wedge states, and in the second step we made a change of variables into 
the total width of the cylinder, $L=\alpha_1+\alpha_2$, and an 
angular parameter $\theta=\alpha_1/L$ describing the distance between 
the $\eta Q\zeta$ and $cV$ insertions on the cylinder. In the third step 
we made an $\mathcal{L}^-$ reparameterization which scales the cylinder 
to unit circumference, and performed the integral over $L$. Now express 
the trace as a correlation function on the cylinder of unit circumference:
\begin{eqnarray}X_4 \lineup = -\int_0^1 d\theta \half 
\mathrm{tr}(\sigma_3\sigma_2\sigma_3\sigma_3\sigma_2\sigma_3\sigma_3\sigma_3)
\left\langle\frac{1}{2}\eta\gamma^{-1}c\d c(1)\ \frac{1}{2}
\gamma^{-1}c\d c(1-\theta)\  B \right\rangle_{C_1},\nonumber\\
\lineup = -\frac{1}{4}\int_0^1 d\theta \Big\langle\big[\eta\gamma^{-1}(1)
\gamma^{-1}(1-\theta)\big]\big[c\d c(1) c\d c(1-\theta)B\big]\Big\rangle_{C_1}.
\end{eqnarray}
Consulting the correlation functions \eq{gginv} and \eq{bcenergy} in appendix
\ref{app:correlators}, this gives
\begin{eqnarray}
X_4\lineup = -\frac{1}{2\pi}\int_0^1 d\theta\, \big(\sin \pi\theta - \pi 
\theta\cos\pi \theta\big),\nonumber\\
\lineup = -\frac{1}{2\pi}\int_0^1 d\theta\, \left(2\sin\pi\theta 
- \frac{d}{d\theta}\theta\sin\pi\theta\right),\nonumber\\
\lineup = -\frac{1}{2\pi}
\left.\left(-\frac{2}{\pi}\cos\pi\theta -\theta\sin\pi\theta\right)\right|_0^1,
\nonumber\\
\lineup = -\frac{2}{\pi^2}.
\end{eqnarray}

Next compute $X_2$. We can simplify the expression as follows: 
\begin{eqnarray}
X_2 \lineup = \Tr\left[\eta Q\zeta\frac{B}{1+K}V\frac{1}{1+K}c\frac{1}{1+K}
\right]+\Tr\left[\eta Q\zeta\frac{B}{1+K}c\frac{1}{1+K}V\frac{1}{1+K}\right],
\nonumber\\
\lineup = \Tr\left[\eta Q\zeta\frac{B}{1+K}Q\zeta\frac{B}{1+K}c\frac{1}{1+K}
\right]+\Tr\left[[B,Q\zeta]\frac{1}{1+K}\eta Q\zeta\frac{B}{1+K}c
\frac{1}{1+K}\right],\nonumber\\
\lineup = \Tr\left[\eta Q\zeta\frac{B}{1+K}Q\zeta\frac{B}{1+K}c\frac{1}{1+K}
\right]+\Tr\left[Q\zeta\frac{B}{1+K}\eta Q\zeta\frac{B}{1+K}c
\frac{1}{1+K}\right]\nonumber\\
\lineup\ \ \ \ \ \ \ \ \ \ \ \ 
-\Tr\left[Q\zeta\frac{1}{1+K}\eta Q\zeta\frac{B}{1+K}c
\frac{B}{1+K}\right],\nonumber\\
\lineup = \Tr\left[\eta \left(Q\zeta\frac{B}{1+K}Q\zeta\frac{B}{1+K}c
\frac{1}{1+K}\right)\right]-\Tr\left[Q\zeta\frac{1}{1+K}\eta Q\zeta
\frac{B}{(1+K)^2}\right],\nonumber\\
\lineup = -\Tr\left[\eta Q\zeta\frac{1}{1+K} Q\zeta
\frac{B}{(1+K)^2}\right],\nonumber\\
\lineup = -\Tr\left[\eta Q\zeta\frac{1}{1+K}cV\frac{B}{(1+K)^2}\right].
\label{eq:X2new}
\end{eqnarray}
Then expanding this out in terms of wedge states as in \eq{expand} we find
\begin{equation}X_2 = -2\int_0^1 d\theta\, (1-\theta)\Tr\Big[(\eta Q\zeta) 
\Omega^{\theta}(cV)B \Omega^{1-\theta}\Big].
\end{equation}
Computing the correlator this becomes
\begin{equation}X_2 = -\frac{1}{\pi}\int_0^1 d\theta\,(1-\theta) 
\big(\sin \pi\theta - \pi \theta\cos\pi \theta\big).
\end{equation}
The second term vanishes by symmetry $\theta\to 1-\theta$. For the first 
term substituting $\theta\to 1-\theta$ simplifies the integral to 
\begin{equation}X_2 = -\frac{1}{\pi}\int_0^1 d\theta\,\theta 
\sin \pi\theta = -\frac{1}{\pi^2}.
\end{equation}
Therefore the coefficients $X_1,...,X_4$ take the values described in \eq{Xs}.

\end{appendix}

\end{document}